\documentclass[aps,prl,reprint,twocolumn,citeautoscript,superscriptaddress,showkeys,showpacs]{revtex4-1}  
\usepackage{xr}
\usepackage[fleqn,tbtags]{amsmath}
\usepackage{amsfonts, amssymb,amsxtra,textcomp}
\usepackage[]{graphicx}
\usepackage{grffile}
\usepackage{bbm}
\usepackage{bm}
\usepackage{color}
\usepackage{graphicx}
\usepackage[colorlinks=true,citecolor=blue,linkcolor=blue]{hyperref}

\allowdisplaybreaks
\usepackage{pdfpages}

\usepackage{color}

\newcommand{\curC}{{\cal C}}

\newcommand{\be}{\begin{equation}}
\newcommand{\ee}{\end{equation}}
\newcommand{\bea}{\begin{eqnarray}}
\newcommand{\eea}{\end{eqnarray}}
\newcommand{\bse}{\begin{subequations}}
\newcommand{\ese}{\end{subequations}}

\usepackage{color}                                      
\definecolor{d_red}{cmyk}{0.00, 0.81, 1.00, 0.27}
\definecolor{d_orange}{cmyk}{0.00, 0.33, 1.00, 0.00}
\definecolor{d_blue}{cmyk}{0.78, 0.47, 0.00, 0.20}
\definecolor{d_lgreen}{cmyk}{0.07, 0.00, 0.79, 0.29}
\definecolor{d_green}{cmyk}{0.66, 0.00, 0.71, 0.56}
\definecolor{d_blue}{cmyk}{0.78, 0.47, 0.00, 0.20}
\definecolor{d_dblue}{cmyk}{0.91, 0.79, 0.00, 0.22}
\definecolor{d_pink}{cmyk}{0.0, 0.79, 0.37, 0.29}
\definecolor{d_purple}{cmyk}{0.16, 0.54, 0.00, 0.70}
\definecolor{d_paleblue}{cmyk}{0.669, 0.338, 0.00, 0.373}
\definecolor{d_dpaleblue}{cmyk}{0.441, 0.290, 0.00, 0.580}
\definecolor{d_brown}{cmyk}{0.0, 0.490, 0.930, 0.350}
\definecolor{d_turquoise}{cmyk}{0.630, 0.04, 0.0, 0.440}

\newcommand{\bfjj}{{\boldsymbol{J}}}
\newcommand{\bfkk}{{\boldsymbol{K}}}

\newcommand{\bfrr}{{\boldsymbol{R}}}

\newcommand{\bfa}{{\boldsymbol{a}}}

\newcommand{\bfh}{{\boldsymbol{h}}}

\newcommand{\bfk}{{\boldsymbol{k}}}

\newcommand{\bfq}{{\boldsymbol{q}}}
\newcommand{\bfr}{{\boldsymbol{r}}}

\newcommand{\bfu}{{\boldsymbol{u}}}

\def\bmx{\begin{pmatrix}}
\def\emx{\end{pmatrix}}

\usepackage{hyperref}
\usepackage[figure,table]{hypcap}               
\hypersetup{
pdftitle = {},
pdfsubject = {},
pdfauthor = {},
pdfkeywords = {},
pdfcreator = {},
pdfproducer = {LaTeX with hyperref},
colorlinks = true,
linkcolor = blue, 
anchorcolor = blue,
citecolor = blue, 
filecolor = red, 
menucolor = red, 
pagecolor = red, 
urlcolor  = blue, 
breaklinks = true,
pdfstartview = FitV,
pdfhighlight = /I,
pdfpagelayout = OneColumn,
hypertexnames=true 
}

\begin{document}

\title{Universal collisionless transport of graphene}

\author{Julia M. Link}
\affiliation{Institute for Theory of Condensed Matter, Karlsruhe Institute of
Technology (KIT), 76131 Karlsruhe, Germany}

\author{Peter P. Orth}
\affiliation{Institute for Theory of Condensed Matter, Karlsruhe Institute of Technology (KIT), 76131 Karlsruhe, Germany}
\affiliation{School of Physics and Astronomy, University of Minnesota, Minneapolis, Minnesota 55455, USA}

\author{Daniel E. Sheehy}
\affiliation{Department of Physics and Astronomy, Louisiana State University,
Baton Rouge, LA, 70803, USA}

\author{J\"org Schmalian}
\affiliation{Institute for Theory of Condensed Matter, Karlsruhe Institute of
Technology (KIT), 76131 Karlsruhe, Germany}
\affiliation{Institute for Solid State Physics, Karlsruhe Institute of Technology (KIT), 76131 Karlsruhe, Germany}

\date{\today}
\begin{abstract}  The impact of the electron-electron Coulomb
interaction on the optical conductivity of graphene has led to a
controversy that calls into question the universality of collisionless transport 
in this and other Dirac materials. Using a lattice calculation that avoids divergences present in previous nodal Dirac
approaches, our work settles this controversy and obtains results in
quantitative agreement with experiment over a wide frequency range. We
also demonstrate that dimensional regularization methods agree, as
long as the scaling properties of the conductivity and the regularization of the
theory in modified dimension are correctly implemented. Tight-binding lattice and nodal Dirac theory calculations
are shown to coincide at low energies even when the non-zero size of the atomic orbital wave function is included, conclusively demonstrating the universality
of the optical conductivity of graphene.
\end{abstract}

\maketitle

In graphene, numerous electronic properties with energy sufficiently
below the scale $ v \Lambda \simeq 1 - 1.5 \mathrm{\, eV}$ are governed by the linear Dirac spectrum with velocity $v$~\cite{RevModPhys.81.109}. Examples are the minimal conductivity in disordered samples~\cite{Novoselov-Nature-2005}, the odd-integer quantum Hall effect at high magnetic fields~\cite{ZhangKim-IQHEGraphene-Nature-2005}, and the observation of Klein tunneling through potential barriers~\cite{PhysRevLett.102.026807}. These observations are explained in terms of non-interacting Dirac fermions, while the electron-electron Coulomb interaction clearly affects other experimental results such as the fractional quantum Hall effect~\cite{DuAndrei-FQHEGraphene-Nature-2009, BolotinKim-FQHEGraphene-Nature-2009}, and the logarithmically enhanced velocity, as seen in magneto-oscillation~\cite{EliasGeim-DiracConesReshapedByInteractionEffects-NatPhys-2010}, angular resolved photoemission spectroscopy~\cite{Siegel12072011} and capacitance measurements of the density of states~\cite{Yu26022013}. 

Given this success, it is remarkable that there exists a rather long-standing controversy in the theoretical description of Coulomb interaction corrections to the optical absorption of graphene~\cite{PhysRevLett.100.046403,0295-5075-83-1-17005, PhysRevB.80.193411,PhysRevB.82.235402, PhysRevLett.110.066602, PhysRevB.86.115408,0295-5075-104-2-27002, PhysRevB.84.045429,PhysRevB.81.245424}. Experiments report an optical transmission close to $97.7\%$~\cite{Nair06062008}, a value that corresponds to non-interacting Dirac electrons. Considering Coulomb interactions within a renormalization group analysis, one finds for the optical conductivity ($\omega\ll v \Lambda$):
\begin{equation}
\label{eq:conduct}
\sigma\left(\omega\right)=\sigma_{0}\left(1+\mathcal{C}\alpha\left(\omega\right)+\cdots\right)\,.
\end{equation}
Here, $\sigma_{0}= \pi e^{2}/(2h)$ is the universal value of the optical conductivity of non-interacting Dirac particles~\cite{PhysRevB.50.7526} and $\alpha\left(\omega\right)=\alpha/[1+\frac{1}{4}\alpha\ln (v \Lambda/\omega)]$ is a running, renormalized, dimensionless coupling constant that measures the strength of the Coulomb interaction at the frequency scale $\omega$, with bare value $\alpha=e^{2}/(\hbar v\epsilon)$~\cite{Gonzalez1994595, PhysRevLett.99.226803}. Here, $e$
is the electron charge and $\epsilon=\left(\epsilon_{1}+\epsilon_{2}\right)/2$ is determined by the dielectric constants $\epsilon_{1,2}$ of the material above and below the graphene sheet.
\begin{figure}[b!]
 \centering 
\includegraphics[width=.45\linewidth]{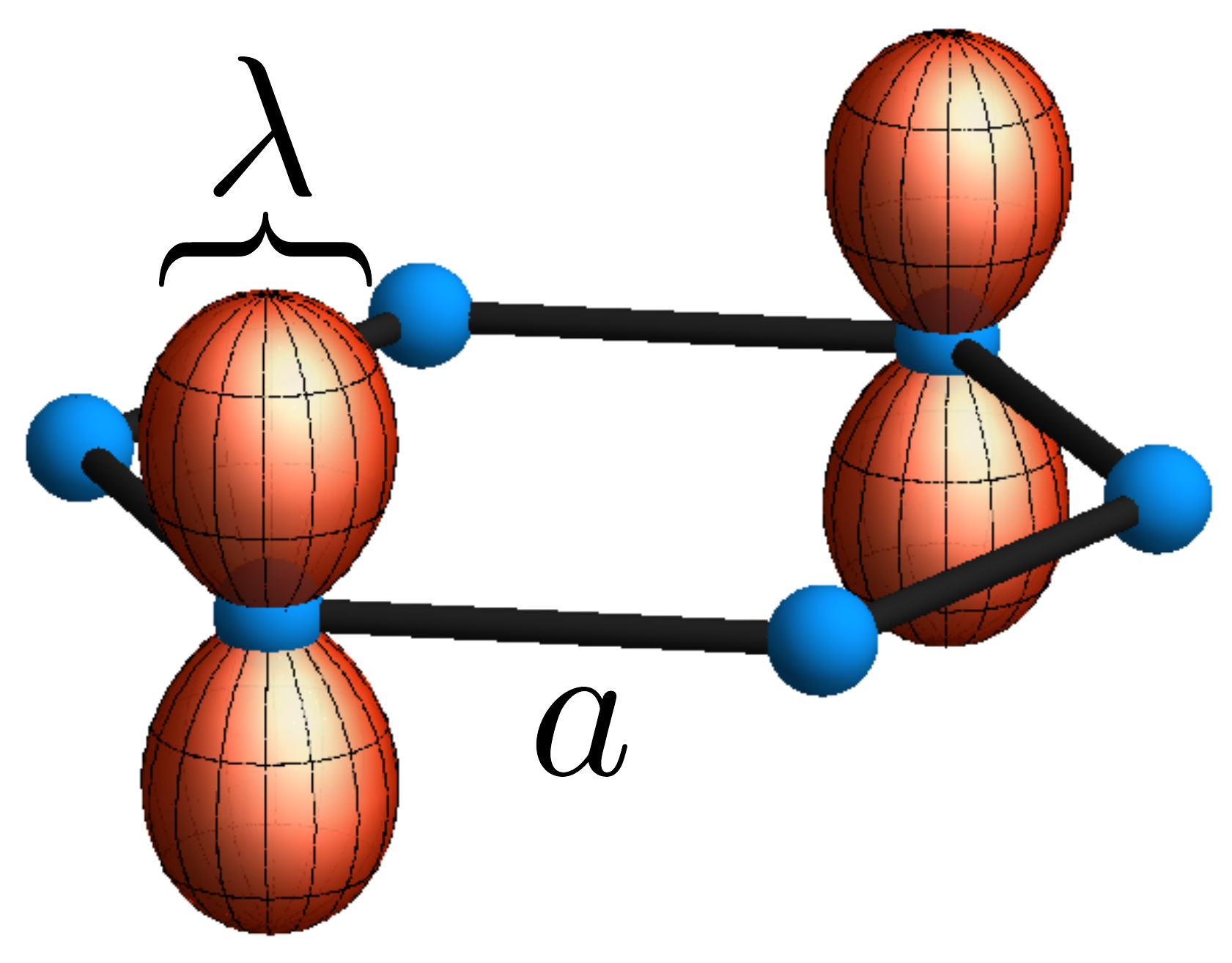}
\caption{(Color online) One plaquette of graphene's honeycomb lattice with blue spheres representing carbon atoms. Carbon-carbon distance is $a$ and two electron $p_z$ orbitals of typical width $\lambda$ are illustrated.}
\label{fig:one} 
\end{figure}

The value of the coefficient $\mathcal{C}$ is the issue of the controversy, with different theoretical approaches yielding different values for ${\cal C}$. The origin of these discrepancies can be traced to the low energy nodal Dirac approximation (NA) for graphene with linear spectrum $\varepsilon\left(\mathbf{q}\right)=\pm v |\bfq|$ for $|\bfq| \leq \Lambda$. A perturbative analysis of corrections due to Coulomb interactions to $\sigma(\omega)$ yields individual Feynman diagrams that are {\em logarithmically divergent\/} in the cutoff $\Lambda$. While these divergences cancel if one adds up all diagrams, the finite result, which determines ${\cal C}$, turns out to be different for different
approaches to handle the divergences. Since $\sigma\left(\omega\right)$
determines the transmission coefficient $T\left(\omega\right)=\left(1+2\pi\sigma\left(\omega\right)/c\right)^{-2}$~\cite{PhysRevB.78.085432}, this issue is experimentally relevant and only a rather small value of $\curC$ is consistent with current observations~\cite{PhysRevB.80.193411}. These controversies were believed to be resolved when two of us demonstrated that a calculation that respects conservation of the electric charge leads to~\cite{PhysRevB.80.193411} %
\begin{align}
\label{Eq:curcdef}
\mathcal{C}&=\frac{19-6\pi}{12}\,,
\end{align}
a value that was first determined by Mishchenko~\cite{0295-5075-83-1-17005}. The essential claim of Ref.~\cite{PhysRevB.80.193411} was that, while different results can be obtained within the NA (as found in earlier work~\cite{PhysRevLett.100.046403}), this ambiguity is eliminated when the Ward identity is enforced.

However, subsequent investigations~\cite{PhysRevB.82.235402,PhysRevLett.110.066602} led to an alternate result for ${\cal C}$, calling into question this picture. In particular, Juricic \emph{et al.}~\cite{PhysRevB.82.235402} used the NA along with dimensional regularization of the integrals (altering the spatial dimension to $d=2-\epsilon$ with $\epsilon \rightarrow 0$ at the end of the calculation), obtaining a much larger value ${\cal C}'=(22-6\pi)/12$ within a calculation that also obeyed the Ward identity at least for finite $\epsilon$. This larger value was also obtained in Ref.~\cite{PhysRevLett.110.066602}, who claimed to perform a tight-binding calculation. Incidentally, using $\mathcal{C'}$ in Eq.~\eqref{eq:conduct} yields results in disagreement with experiment. The authors of Ref.~\cite{PhysRevLett.110.066602} concluded that the source of the error was the linearized spectrum and concluded that a proper treatment of the spectrum in the entire Brillouin zone (BZ) is needed to determine the optical conductivity. It was added that this unexpected behavior is related to a chiral anomaly or due to non-local optical effects~\cite{PhysRevB.90.045137}.

Given these discrepancies, an obvious question is whether ${\cal C}$ is indeed a universal number. If states in the entire BZ matter, one could easily construct new dimensionless quantities $\gamma$ and the coefficient ${\cal C}$ in Eq.~\eqref{eq:conduct} might depend on $\gamma$. Then, distinct analytic results would merely correspond to different limits of ${\cal C}\left(\gamma\right)$. An example for such a dimensionless quantity is $\gamma=\lambda/a$, where $a\approx 1.42\,\text{\AA}$ is the carbon-carbon distance and $\lambda$ the size of the $p_z$-orbital Wannier function of the $sp^2$ hybridized graphene lattice (see Fig.~\ref{fig:one}). Then, only a detailed quantum chemical analysis would be able to determine the correct optical conductivity, even for frequencies small compared to the bandwidth. This would imply the breakdown of the widely-used NA for graphene.

In this paper we start from a lattice tight-binding description of
graphene and determine the optical conductivity, including leading
Coulomb corrections, in the collisionless regime. Allowing for a finite
extent of the Wannier functions $\lambda$, we demonstrate that the constant ${\cal C}$ is indeed universal, i.e. independent of the ratio $\lambda/a$, and takes a value that is, within the numerical accuracy, given in Eq.~(\ref{Eq:curcdef}). We explain why previous lattice based attempts~\cite{PhysRevLett.110.066602} failed to reach the correct conclusion and show how to obtain Eq.~(\ref{Eq:curcdef}) even in case of the dimensional regularization scheme used in Ref.~\cite{PhysRevB.82.235402}. The latter conclusion is fully consistent with the recent field theory analysis of Ref.~\cite{0295-5075-107-5-57001}. Thus, our work finally settles all aspects of this controversy. It demonstrates that the Dirac cone approximation can be safely applied for low energy properties and that the longitudinal optical conductivity is not affected by a chiral anomaly or states far from the Dirac cone and that no subtlety due to non-local effects in the conductivity occurs. Finally, it shows how to properly include interaction corrections within the lattice theory, which is essential for physical quantities where a NA cannot be applied.

Graphene is a honeycomb lattice of carbon atoms spanned by the (triangular) Bravais lattice vectors $\boldsymbol{R}_i = i_{1} \bfa_{1}+ i_{2} \bfa_{2}$ with $i_{1,2} \in \mathbb{Z}$ and $\bfa_{1,2} = \frac{\sqrt{3}}{2} a \bigl(\pm 1, \sqrt{3} \bigr)$, and basis vectors $\bfu_{a,b}$ (one choice is $\bfu_{a}=0$, $\bfu_b = (0, -a)$). We introduce a spinor composed of electron creation operators $c_{\bfrr_i \sigma}^\dag = (a^\dag_{\bfrr_{i}\sigma}, b^\dag_{\bfrr_{i}\sigma})$ which create an electron with spin $\sigma$ on the corresponding lattice site $(\bfrr_i, l)$ with $l=a,b$. With these definitions it follows that the tight-binding Hamiltonian reads $H_{0}=\sum_{{\bf k}l\sigma}c_{{\bf k}l\sigma}^{\dagger}{\cal H}_{{\bf k}ll'}c_{{\bf k}l\sigma}$ with $c_{\bfrr_i \sigma} = \frac{1}{\sqrt{N}} \sum_{\bfk} e^{i \bfk \bfrr_i } c_{\bfk \sigma}$. In case of only nearest-neighbor hopping $t$, one finds ${\cal H}_{{\bf k}}=\bfh_{\bfk} \cdot \boldsymbol{\sigma}$ with the vector $\bfh_{\bfk} =(\text{Re} \, h_{\bfk}, \text{Im} \, h_{\bfk} )$ given by $h_{\bfk} =-t \bigl( 1 + e^{-i \bfk \bfa_{1}} + e^{-i \bfk \bfa_{2} } \bigr)$ and Pauli matrices $\boldsymbol{\sigma} = (\sigma_x, \sigma_y)$. A linear Dirac spectrum emerges near ${\bf K}_{\pm}=\frac{2\pi}{3 a} \bigl( \pm \frac{1}{\sqrt{3}}, 1 \bigr)$. The current operator is given by~\cite{Mahan-ManyParticlePhysics} $\bfjj_{\bfrr_i} = -\frac{iet}{\hbar} \sum_{\boldsymbol{\delta}_\alpha} [ (\boldsymbol{\delta}_\alpha + \bfu_b - \bfu_a) b^\dag_{\bfrr_i + \boldsymbol{\delta}_\alpha} a_{\bfrr_i} - \text{h.c.} ]$ with nearest-neighbor Bravais lattice vectors $\boldsymbol{\delta}_\alpha$ ($\alpha = 1,2,3$). 
Electrons interact via the Coulomb interaction
\begin{equation}
\label{eq:5}
H_{int}=\frac{e^{2}}{2}\sum_{\sigma\sigma'}\int d^{3}r \, d^{3}r' \; \frac{\psi_{\bfr \sigma}^{\dagger} \psi_{\bfr' \sigma'}^{\dagger} \psi_{\bfr' \sigma'} \psi_{\bfr \sigma}}{\epsilon |\bfr -\bfr'|} \,, 
\end{equation}
where the field operators $\psi_{\bfr \sigma} = \sum_{\bfrr_{i} l}\varphi(\bfr - \bfrr_{i}- \bfu_{l}) c_{\bfrr_i l \sigma}$ are defined via the Wannier $p_z$-atomic orbitals $\varphi(\bfr)$ localized on the $sp^2$-hybridized carbon atom at site $(\bfrr_i, l)$. 
\begin{figure}
\includegraphics[width=.72\linewidth]{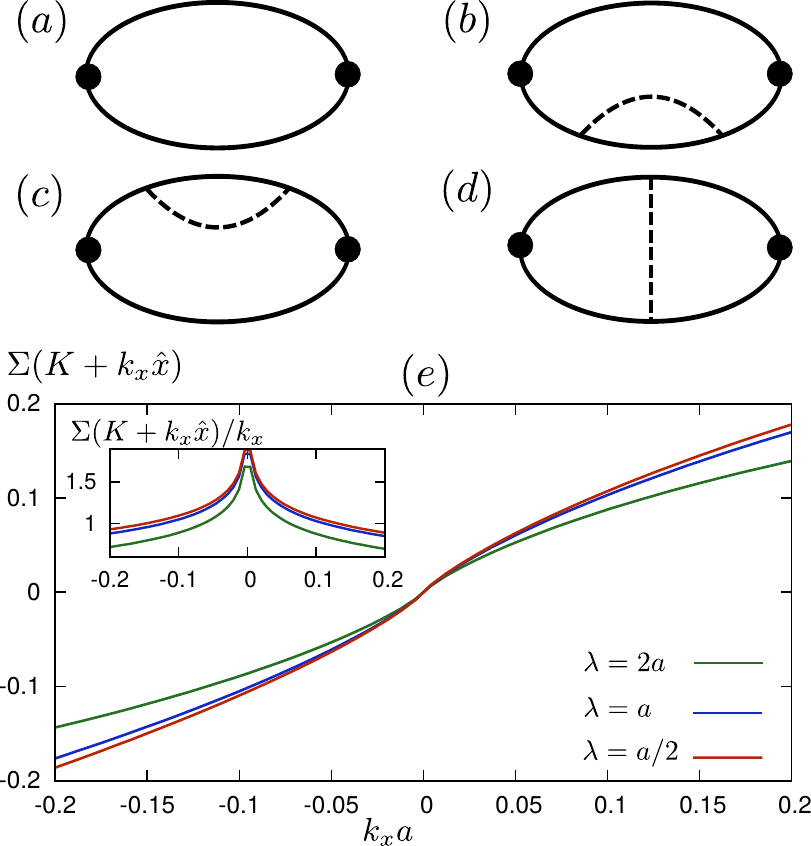} 
\caption{(Color online) Panel $(a)$ shows the Feynman diagram for the non-interacting current-current correlation function $\chi^{(0)}_J$. Panels $(b)-(d)$ show the lowest order Coulomb interaction corrections: $(b)-(c)$ are self-energy diagrams and $(d)$ the vertex correction. Panel $(e)$ shows the lattice self-energy $\Sigma(\boldsymbol{k})$ close to the Dirac node $K\equiv K_+$ for different Wannier orbital sizes $\lambda/a$; inset shows logarithmic divergence of the slope  (= velocity correction) close to the node. }
\label{fig:feynman} 
\end{figure}
In the evaluation of the Coulomb matrix elements, we assume that $\varphi^{*} (\bfr-\bfrr_{i}- \bfu_{l} ) \varphi( \bfr - \bfrr_{j} -\bfu_{m})$ is small unless $i=j$ and $l=m$ such that we obtain in momentum space (for details see the supplemental material (SM)~\cite{Supplemental-OptCondGraphene})
\begin{align}
H_{int} & = \frac{1}{2} \int \frac{d^{2}q}{(2\pi)^{2}} V(\bfq) \sum_{l l'} e^{i \bfq (\bfu_{l}-\bfu_{l'})} \nonumber \\ 
& \times \sum_{\bfk \bfk' \sigma\sigma'} c_{\bfk +\bfq l \sigma}^{\dagger} c_{\bfk'-\bfq l'\sigma'}^{\dagger}c_{\bfk' l' \sigma'}c_{\bfk l \sigma} 
\label{eq:Hlattice}
\end{align} with $V (\bfq) = 4 \pi e^2 \int_{-\infty}^{\infty}
\frac{dq_{z}}{2\pi} \frac{ |\rho(\bfq, q_{z})|^{2}}{\epsilon (q^{2} +
q_{z}^{2})}$ determined by the electron density of the
three-dimensional atomic orbital $\rho\left({\bf q},q_{z}\right)=\int
d^{3}r\left|\varphi(\bfr )\right|^{2}e^{i(\bfq \bfr_{\parallel}+q_{z}
z)}$, where $\bfr_{\parallel}$ is the projection of $\bfr$ into the
graphene plane. Using the $2p_{z}$-orbitals with effective Bohr radius
$a_{B}^{*}$, we obtain $V(\bfq)=2\pi e^{2}\mathcal{F}(\bfq)/(\epsilon |\bfq|)$, where the
form factor was fitted to $\mathcal{F}(\bfq) = \exp(- |\bfq|
a_{B}^{*})$ and $a_{B}^{*} \simeq 0.9 \text{\AA}$~\cite{Groenqvist-EPJB-2012} . In the following,
we use $V(\bfq) = 2 \pi e^2 \exp(-|\bfq|^2 \lambda^2/2)/|\bfq|$ that follows from
a Gaussian wavefunction, with $\lambda$ corresponding to the size of the orbital (see
Fig.~\ref{fig:one}). 

All momentum vectors in Eq.~(\ref{eq:Hlattice}) are two-dimensional. Crucial for our subsequent analysis is the fact that the sums $\sum_{\bfk,\bfk'}$ in Eq.~(\ref{eq:Hlattice}) run over the first BZ, while the integral over $\bfq$ goes over the infinite momentum space, i.e., it is a combined sum over transferred momenta of the BZ and a sum over all reciprocal lattice vectors, a distinction that was ignored in earlier work~\cite{PhysRevLett.110.066602}. This follows from the fact that the electron density of the orbitals $|\varphi(\bfr)|^{2}$ is not confined to the discrete lattice points.

We determine the real part of the optical conductivity via the Kubo formula 
\begin{equation}
\sigma\left(\omega\right)=-\frac{\mathrm{Im}\chi_{J}\left(\omega\right)}{\omega} \,,
\end{equation}
where $\chi_{J}(\omega)$ is the retarded current-current response function. Expanding perturbatively in orders of the Coulomb interaction strength $\alpha$ gives $\chi_{J}=\chi_{J}^{\left(0\right)}+\chi_{J}^{\left(1\right)}+\ldots$, where $\chi_{J}^{(0)}$ refers to non-interacting electrons (see diagram (a) in Fig.~\ref{fig:feynman}). We evaluate $\chi_{J}^{(0)}$ by first analytically continuing $i \omega \rightarrow \omega + i \delta$ and then numerically computing the remaining one-dimensional integral. Beyond the Dirac approximation our results differ from previously reported ones~\cite{Nair06062008,PhysRevB.78.085432} which has consequences for the experimentally observable optical transmission through graphene (see~\cite{Supplemental-OptCondGraphene} for details and also Fig.~\ref{fig:4}). Then, $\chi_{J}^{(1)}$ is the leading order interaction correction depicted in Fig.~\ref{fig:feynman}(b-d) with self-energy ($b$,$c$) and vertex ($d$) parts
\begin{widetext}
\begin{align}
\label{eq:4}
\chi_{J}^{(1,bc)}(i\omega) & = - T^{2} \sum_{\bfk \epsilon \epsilon' \mu}\int\frac{d^{2}q}{(2\pi)^{2}} V(\bfq ) \text{Tr} \bigl( J_{\bfk\mu} G_{\bfk,i\omega+i\epsilon}J_{\bfk\mu}G_{\bfk,i\epsilon} M_{\bfq}G_{\bfk+\bfq,i\epsilon'} M_{-\bfq}G_{\bfk,i\epsilon} \bigr)\\
\label{eq:3}
\chi_{J}^{(1,d)} (i\omega) & = \frac{T^{2}}{2} \sum_{\bfk \epsilon \epsilon' \mu} \int\frac{d^{2}q}{(2\pi)^{2}} V(\bfq) \text{Tr} \bigl( J_{\bfk\mu} G_{\bfk,i\omega+i\epsilon} M_{\bfq} G_{\bfk+\bfq,i\omega+i\epsilon'} J_{\bfk\mu} G_{\bfk+\bfq,i\epsilon'} M_{-\bfq} G_{\bfk,i\epsilon} \bigr) \,.
\end{align}
\end{widetext}
Here, $G_{\bfk,i\omega}=(i\omega-{\cal H}_{\bfk})^{-1}$ denotes the bare Green's function and the matrix $M(\bfq) = \left( \begin{smallmatrix} \exp(i \bfq \bfu_a) & 0 \\ 0 & \exp(i \bfq \bfu_b) \end{smallmatrix} \right)$ accounts for the spatial separation of the two carbon basis atoms (see Eq.~\eqref{eq:Hlattice}). It plays an important role in the following evaluation of $\chi^{(1)}_J$ as it renders the integration over momentum $\bfq$ finite. To see this explicitly, let us analyze $\chi^{(1,bc)}_J$ and define a self-energy as
\begin{align}
\label{eq:finalsigma}
\Sigma(\bfk) = - \int \frac{d^2q}{(2\pi)^2} V(\bfq) \sum_\epsilon M_{-\bfq} G_{\bfk+\bfq, i \epsilon} M_\bfq \,.
\end{align} 
Upon evaluating the frequency integration, we obtain $\Sigma(\bfk)= \left( \begin{smallmatrix}0 & \Sigma_{12}\\
\Sigma_{12}^{*} & 0
\end{smallmatrix} \right)$ with 
\begin{align} 
\label{Eq:selfh} 
\Sigma_{12}(\bfk) = -\frac{1}{2}
\int \frac{d^2q}{(2\pi)^2}V(\bfq) e^{i\phi(\bfk+\bfq)} e^{i\bfq (\bfu_b - \bfu_a)}\,,
\end{align}
 where $\exp[i\phi(\bfk)]=h_{\bfk}/|h_{\bfk}|$. Since the remaining summations to obtain $\chi^{(1,bc)}_J$ are restricted to the first BZ (see Eq.~\eqref{eq:4}), any ultraviolet (UV) divergences in this contribution, such as found in the NA, must come from Eq.~(\ref{Eq:selfh}). In the NA to $\Sigma_{12}(\bfk)$, one finds for $\bfk$ near node $\bfkk_\pm$ that $\Sigma_{12}(\bfkk_\pm + \bfq) \sim (\mp q_x + i q_y ) \ln(\Lambda/|\bfq|)$ with $\Lambda$ a cutoff introduced to regularize the UV divergence. Here, we will not introduce any such cutoff, since the integral in Eq.~(\ref{Eq:selfh}) is convergent due to the regularizing effect of a finite lattice constant $a$. Even for atomic orbitals of zero width ($\lambda = 0$), the self-energy remains finite due to the factor ${\rm e}^{i\bfq (\bfu_b - \bfu_a)}$ in Eq.~\eqref{Eq:selfh} that oscillates rapidly at large $q$. We explicitly demonstrate this in~\cite{Supplemental-OptCondGraphene} by performing a Fourier transformation of the self-energy to real-space $\Sigma(\bfk) = - \frac{e^2}{2} A \sum_{\bfrr_i} e^{i \bfk \bfrr_i} \frac{F(\bfrr_i)}{|\bfrr_i + \bfu_b - \bfu_a|}$, where $A$ is the area of the graphene sheet, $F(0) \neq 0$ and $|F(\bfrr_i)|$ decays sufficiently quickly to ensure convergence of the sum. All our real-space summations run over $4.6 \times 10^4$ Bravais lattice vectors $\bfrr_i$ of smallest magnitude~\cite{Supplemental-OptCondGraphene}.  
\begin{figure}
\includegraphics[width=.8\linewidth]{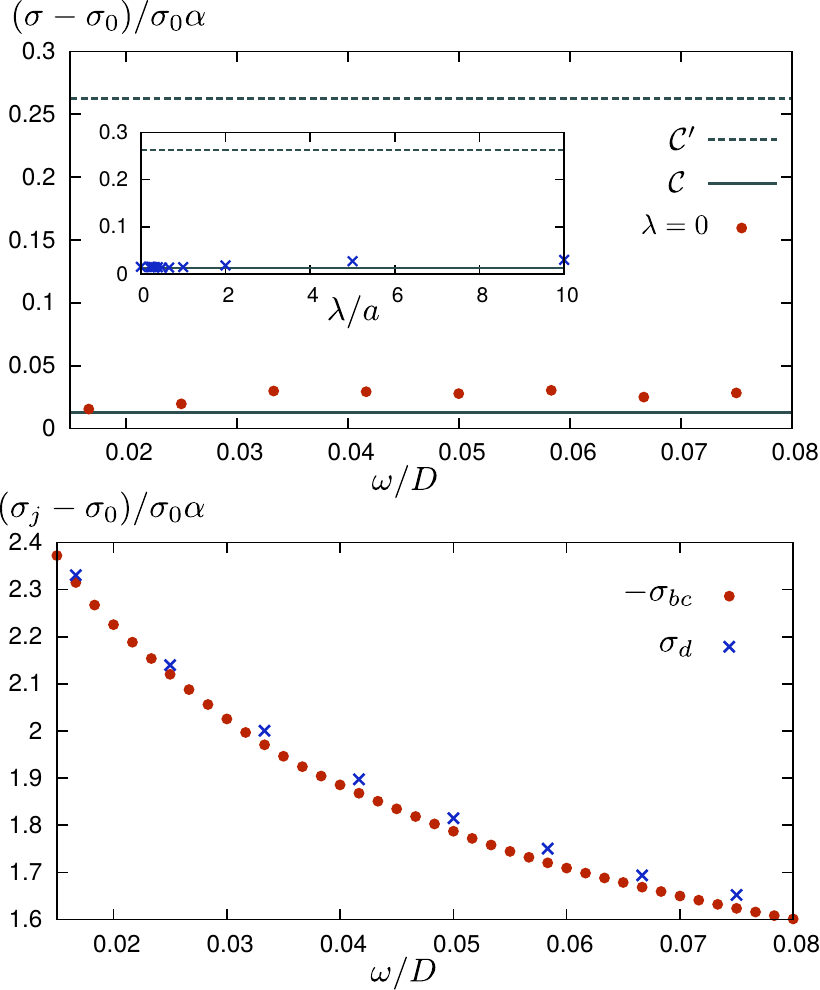} 
\caption{(Color online) The upper panel shows the interaction correction coefficient as determined by our full lattice calculation (red dots) as a function of frequency $\omega/D$. The lattice result is in clear agreement with the predicted value Eq.~(\ref{Eq:curcdef}) from nodal theory~\cite{0295-5075-83-1-17005,PhysRevB.80.193411}. The inset shows that the result in the low-frequency limit ($\omega/D = 0.015$) is independent of the ratio $\lambda/a$ (i.e., universal).
 Lower panel shows individual contributions of self-energy $\sigma_{bc}$ and vertex correction $\sigma_d$ diagrams.  }
\label{fig:correctioncoeff}
\end{figure}

In Fig.~\ref{fig:feynman}(b), we show our numerical result for $\Sigma_{12}(\bfk)$ for $\bfk$ near the node, showing the logarithmic divergence of the
{\em slope} as the node is approached. This is a well known property found in the Dirac approximation that we now see holds in the full tight-binding theory as well. To obtain $\chi^{(1,bc)}_{J}$, we insert our result for the self energy into Eq.~\eqref{eq:4}, analytically continue $i\omega\to\omega+i\delta$, and then evaluate the remaining integral over $\bfk$, which is clearly convergent as it is restricted to the first BZ. 

The vertex contribution $\chi_J^{(1,d)}$ is evaluated in a similar way (see~\cite{Supplemental-OptCondGraphene} for details). The presence of the matrix $M(\bfq)$ inside the trace again ensures convergence of the $\bfq$-integration. The optical conductivity $\sigma$ and the interaction correction coefficient $\mathcal{C}$ are then determined by adding all contributions as $\sigma = \sigma^{(0)} +\sigma^{(1)} + \ldots$ with $\sigma^{(i)} = - \text{Im} \chi^{(i)}_J/\omega$. As shown in Fig.~\ref{fig:correctioncoeff}, the individual contributions to $\sigma^{(1)} =  \sigma_0 \alpha (\sigma_{bc} + \sigma_{d})$ diverge logarithmically in the low frequency limit $\omega/D$, where $D = 6 t$ is the bandwidth. Their sum, however, 
remains finite and yields (within numerical accuracy) the coeffient $\mathcal{C}$ in Eq.~\eqref{Eq:curcdef}, independently of the ratio $\lambda/a$, demonstrating
the universal nature of the optical conductivity and transparency of graphene (see Fig.~\ref{fig:4}).  
\begin{figure}[t]
  \centering
  \includegraphics[width=.85\linewidth]{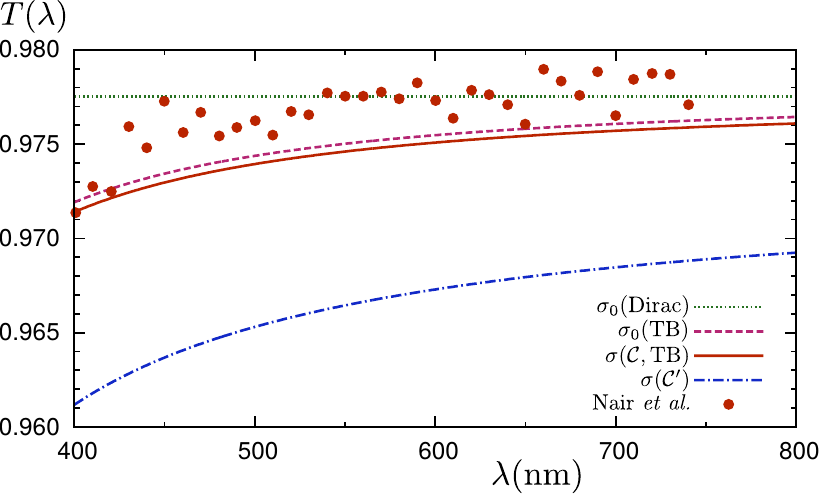}
  \caption{(Color online) Optical transmission through graphene as function of wavelength $\lambda$. Comparison of experimental results (red dots) from Ref.~\onlinecite{Nair06062008} and theory: non-interacting Dirac fermions $\sigma_0(\text{Dirac})$ and tight-binding theory $\sigma_0(\omega, \text{TB})$; interacting tight-binding prediction $\sigma(\mathcal{C} , \text{TB}) = \sigma_0(\omega,\text{TB}) [1 + \mathcal{C} \alpha(\omega)]$, and $\sigma(\mathcal{C}') = \sigma_0(\omega,\text{TB}) [1 + \mathcal{C}' \alpha(\omega)]$. Note that $\sigma_0(\omega, \text{TB})$ deviates from $\sigma_0(\text{Dirac})$ more strongly than previously reported in Refs.~\onlinecite{Nair06062008,PhysRevB.78.085432}. }
  \label{fig:4}
\end{figure}

How does our numerical result of ${\cal C}$ given in Eq.(2) compare
with Ref.~\onlinecite{PhysRevLett.110.066602}, who claim to have
performed an evaluation of the conductivity of the tight-binding
model, but find the larger value ${\cal C}'$? Following the details of
Ref.~\onlinecite{PhysRevLett.110.066602} included in the supplementary
material of that paper we find that, in the end, the authors do not evaluate the
conductivity numerically, but perform a nodal approximation and
regularize diverging integrals in a fashion that violates charge
conservation. The final expression of the conductvity coefficient of
Ref.~\onlinecite{PhysRevLett.110.066602} is not the correct lattice
version of the conductivity anyway, as it lacks the distinction between
BZ restricted and unrestricted momentum integrations, discussed
above. 

Having established within a tight-binding model that the
frequency-dependent conductivity is proportional to the coefficient
$\curC$, next we turn to the question of why results based on
dimensional regularization (DR) (as presented in the detailed calculations
of Juricic \emph{et al.}~\cite{PhysRevB.82.235402}) seemingly yield a
different result. In fact, this issue has already been understood by Teber and
Kotikov (TK) within a field theoretic approach for Dirac fermions in
$d=2-\epsilon$ spatial dimensions~\cite{0295-5075-107-5-57001}.  

The TK calculation (which we review in the SM~\cite{Supplemental-OptCondGraphene}), 
shows that properly regularizing the theory of graphene in $d=2-\epsilon$ dimensions
 (using the modified minimal subtraction, or $\overline{MS}$, scheme) 
gives an additional renormalization of the conductivity that finally leads to Eq.~(\ref{eq:conduct}).  
We add to the insight of TK by examining the conductivity
scaling relation in $d=2-\epsilon$ dimensions:
\be
\label{sigmaepsilon}
\sigma(\omega,\alpha) = \lim_{\Lambda\to\infty} b^{\epsilon} \sigma\big(\omega/Z_T,\alpha(b),\Lambda).
\ee
Here, $\Lambda$ is an ultraviolet cutoff in our theory, introduced within the Wilson momentum-shell RG (WRG),
that relates the true conductivity (left side) to the renormalized conductivity.
In fact, as we now show, the WRG leads to an ultraviolet quirk in the evaluation of the right side.  In particular, we find that 
the limit $\Lambda \to \infty$ must be taken {\em after\/} the limit $\epsilon \to 0$ (returning to the physical dimensionality) has been taken. 
If the limit $\Lambda \to \infty$ is taken first, then, although the theory is finite in $2-\epsilon$ dimensions, additional contributions
(coming from the large momentum part of the integral) emerge at  $\epsilon \to 0$.

 This can be seen by directly considering the evaluation of the
conductivity within the density-correlator approach of
Mishchenko~\cite{0295-5075-83-1-17005}, at fixed $\Lambda$ but in
spatial dimension $d = 2-\epsilon$.    There, only one diagram (of the
form of diagrams (b) and (c), with a self-energy subdiagram)
possesses a singular structure making the evaluation rather simple.
With details provided in the SM~\cite{Supplemental-OptCondGraphene},
we find that the final result depends crucially on the order of
limits. The resulting interaction coefficient, in
$d=2-\epsilon$ dimensions and at fixed $\Lambda$, has the form: 
\be
\label{curcfin}
\curC(\epsilon,\frac{\omega}{\Lambda}) = \frac{22-6\pi - 3\big(\frac{\omega}{v\Lambda}\big)^\epsilon}{12}\,,
\ee
showing that, to obtain the result Eq.~(\ref{Eq:curcdef}), one indeed must take the limit $\epsilon \rightarrow 0$ before taking $\Lambda \rightarrow \infty$. The case of $\Lambda \rightarrow \infty $ at fixed $\epsilon$ instead yields $\curC'$. However, since $\epsilon = 0$ is the marginal dimension, additional singularities may appear when the limit $\epsilon\to 0$ is taken, and the best strategy is to regularize the full theory via the $\overline{MS}$ prescription as done by TK, who find a modification of the bare-bubble contribution (diagram (a) of  Fig.\ref{fig:feynman}) that combines with Eq.~(\ref{curcfin}) to finally arrive at Eq.~(\ref{Eq:curcdef}).

In conclusion, we have evaluated the optical conductivity of graphene including the lowest order Coulomb interaction corrections within a full lattice tight-binding approach. We correct previous results of the non-interacting conductivity beyond the Dirac limit. Considering interactions, we explicitly show that $\sigma$ is universal and independent of other dimensionless quantities such as the ratio of the atomic orbital width to the lattice constant (for frequencies $\omega  < v \Lambda$). 
Our work validates previous Dirac approximation calculations, and resolves a long-standing controversy about the correct way to regularize the Dirac theory. Since descriptions of electronic systems by effective low-energy models like the Dirac Hamiltonian of graphene are the cornerstone of condensed matter physics, it is gratifying that our work confirms the quantitative accuracy of this method.

We gratefully acknowledge useful discussions with U. Briskot, I. Herbut,
A. Mirlin, E. Mishchenko, C. Seiler, S. Teber and O. Vafek. D.E.S. was supported by the National Science Foundation Grant No. DMR-1151717 and by the German
Academic Exchange Service (DAAD). J.M.L. thanks the
Carl-Zeiss-Stiftung for financial support. This work was performed on
the computational resource bwUniCluster funded by the Ministry of
Science, Research and the Arts Baden-W\"urttemberg and the
Universities of the State of Baden-W\"urttemberg, Germany, within the
framework program bwHPC. The Young  Investigator  Group  of
P.P.O. received  financial  support  from  the  ``Concept  for  the
Future'' of the Karlsruhe Institute of Technology (KIT) within the
framework of the German Excellence Initiative.


%

\pagebreak


\section*{Supplementary material}

# Supplementary Material: Universal collisionless transport of graphene

Julia M. Link,[1] Peter P. Orth,[1, 2] Daniel E. Sheehy,[3] and Jörg Schmalian[1, 4]

[1] *Institute for Theory of Condensed Matter, Karlsruhe Institute of Technology (KIT), 76131 Karlsruhe, Germany*
[2] *School of Physics and Astronomy, University of Minnesota, Minneapolis, Minnesota 55455, USA*
[3] *Department of Physics and Astronomy, Louisiana State University, Baton Rouge, LA, 70803, USA*
[4] *Institute for Solid State Physics, Karlsruhe Institute of Technology (KIT), 76131 Karlsruhe, Germany*

(Dated: November 18, 2015)

## CONTENTS

## OVERVIEW

This supplementary material section contains two main sections describing our calculations of the frequency-dependent conductivity of graphene. Although our calculations are relatively straightforward, given the discrepancies in the literature regarding the evaluation of this quantity we felt it was necessary to provide considerable detail.

In Sec. A, we describe our calculation of the conductivity of graphene within the tight-binding approach, including the bare bubble as well as leading-order interaction corrections. As discussed in the main text, these contributions do not possess any of the divergences that arise in the nodal approximation, with the only uncertainties being numerical accuracy. However, as in the nodal approximation, these contributions do involve integrations over the entire momentum space (arising from the Fourier transform of the Coulomb interaction). We demonstrate that these integrations are not divergent, with the convergence being demonstrated by analyzing the corresponding contributions in real-space. As noted in the main text, our final results for the interaction corrections to the conductivity disagree with the findings of Rosenstein et al[9] who claim to perform a calculation "directly in the tight binding approach". In fact, these authors do not perform a tight binding calculation and, instead, make a nodal approximation that leads to divergent contributions. In contrast, our tight binding calculation makes *no nodal approximations*.

In Sec. B, we consider the Dirac theory of graphene and the intreaction contributions whose divergences needs to be regularized. We first elaborate in more detail the calculation of Teber et al.,[7] where the divergent integrals were regularized using dimensional regularization and renormalized using a field theoretical minimal subtraction scheme or in other words a continuum RG. Next, we will show at the example of the Mishchenkov's approach using the density-density correlator, that the combination of dimensional regularization with Wilson momentum-shell RG will yield a UV quirk. We will obtain a non-commuting order of limit between the UV-cutoff $\Lambda$ of the Wilson RG and the

parameter of the dimensional regularization $\epsilon$. Only when the UV-cutoff $\Lambda$ is taken at the very end of the calculation, we obtain when combining WRG and DR the correct correction coefficient

$$\mathcal{C} = \frac{19 - 6\pi}{12} \,. \tag{12}$$

## Appendix A: Tight-binding theory of graphene

In this section we provide details of the tight-binding calculation of the conductivity of graphene. We first review the single-particle Hamiltonian on the honeycomb lattice, and obtain the bare Green's function and the current operator. Subsequently, we consider the Coulomb interaction term in the Hamiltonian (making sure to account for the two-atom basis) before considering the electron self energy. Finally, we review the Kubo formula relating the frequency-dependent conductivity to a current-current correlation function and present our calculations of the contributions to this quantity to zeroth order and first order in perturbation theory.

### 1. Single-particle Hamiltonian

The single-particle Hamiltonian, describing electrons hopping on the honeycomb lattice of graphene, is

$$\mathcal{H}_0 = -t \sum_{\langle i,j \rangle, \sigma} (a_{i\sigma}^{\dagger} b_{j\sigma} + b_{j\sigma}^{\dagger} a_{i\sigma}), \tag{A1}$$

where $i$ and $j$ are nearest-neighbor lattice sites and the $a_{i\sigma}$ and $b_{j\sigma}$ operators defined on the red and blue sublattices, respectively. Henceforth, we shall suppress the spin index $\sigma$. To express this in Fourier space, we first define our Bravais lattice to be the red sublattice of Fig. 1, defined by two primitive Bravais lattice vectors $\mathbf{a}_1$ and $\mathbf{a}_2$:

$$\mathbf{a}_1 = a \big( \frac{\sqrt{3}}{2} \hat{x} + \frac{3}{2} \hat{y} \big), \tag{A2}$$

$$\mathbf{a}_2 = a \big( -\frac{\sqrt{3}}{2} \hat{x} + \frac{3}{2} \hat{y} \big), \tag{A3}$$

where $a$ is the nearest neighbor distance. Then we define the Fourier series

$$a_i = \frac{1}{\sqrt{N}} \sum_{\mathbf{k}} e^{i\mathbf{k}\cdot\mathbf{R}_i} a_{\mathbf{k}}, \tag{A4}$$

$$a_{\mathbf{k}} = \frac{1}{\sqrt{N}} \sum_{i} e^{-i\mathbf{k}\cdot\mathbf{R}_i} a_i, \tag{A5}$$

and similarly for the $b_i$, with $\mathbf{k}$ defined to be in the first Brillouin zone (BZ).

Upon inserting these definitions into $\mathcal{H}_0$, we find the well-known result

$$\mathcal{H}_0 = -t \sum_{\mathbf{k}} \psi^{\dagger}(\mathbf{k}) \begin{pmatrix} 0 & h(\mathbf{k}) \\ h^*(\mathbf{k}) & 0 \end{pmatrix} \psi(\mathbf{k}), \tag{A6}$$

$$h(\mathbf{k}) = 1 + e^{i\mathbf{k}\cdot\mathbf{a}_1} + e^{i\mathbf{k}\cdot\mathbf{a}_2} = 1 + 2\cos\big(\frac{\sqrt{3}}{2} k_x a\big) e^{i\frac{3}{2}k_y a}, \tag{A7}$$

where we defined the spinor

$$\psi(\mathbf{k}) = \begin{pmatrix} a_{\mathbf{k}} \\ b_{\mathbf{k}} \end{pmatrix} . \tag{A8}$$

As is well known, $\mathcal{H}_0$ possesses two independent nodes in the BZ, with fermions near these nodes having an approximately linear dispersion with velocity $v = \frac{3}{2}\frac{ta}{\hbar}$. Below we shall often use units in which $\hbar = t = a = 1$.

The bare Green's function, which we will need for our perturbative calculation of the conductivity, is:

$$G(\mathbf{k}, \omega) = \Big[ i\omega\sigma_0 - \begin{pmatrix} 0 & -th(\mathbf{k}) \\ -th(-\mathbf{k}) & 0 \end{pmatrix} \Big]^{-1} = \begin{pmatrix} i\omega & th(\mathbf{k}) \\ th^*(\mathbf{k}) & i\omega \end{pmatrix}^{-1} . \tag{A9}$$

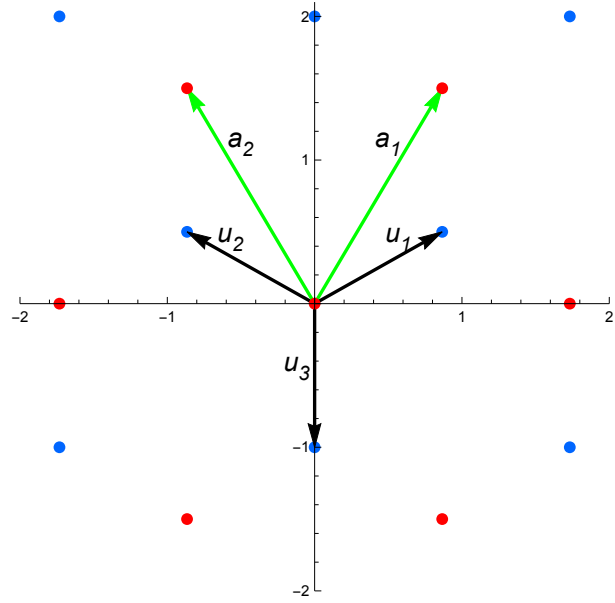

FIG. 1. (Color online) Plot of honeycomb crystal lattice of graphene, showing two interpenetrating (red and blue) sublattices. The Bravais lattice is defined to be the red sublattice, with a two-atom basis defined by $\mathbf{v}_1 = 0$ (the red points) and $\mathbf{v}_2 = \mathbf{u}_3 = -a\hat{y}$ (the blue points).

To define the current operator, we follow Mahan to write the spatially-integrated current density as:

$$\mathbf{J} = i[\mathcal{H}, \mathbf{P}], \tag{A10}$$

with the polarization $\mathbf{P} = \sum_i \mathbf{r}_i n_i$, with $n_i = a_i^\dagger a_i$ on the red sublattice and $n_i = b_i^\dagger b_i$ on the blue sublattice. Here, $\mathbf{r}_i$ are the real-space locations of the lattice sites. Taking into account the fact that the blue sublattice is shifted relative to the red sublattice, we obtain

$$\mathbf{J} = \sum_{\mathbf{k}} \psi^\dagger(\mathbf{k}) \hat{\mathbf{J}}(\mathbf{k}) \psi(\mathbf{k}), \tag{A11}$$

with

$$\hat{\mathbf{J}}(\mathbf{k}) = -it \Big[ \mathbf{u}_3 \begin{pmatrix} 0 & 1 \\ -1 & 0 \end{pmatrix} + \mathbf{u}_1 \begin{pmatrix} 0 & e^{i\mathbf{k}\cdot\mathbf{a}_1} \\ -e^{-i\mathbf{k}\cdot\mathbf{a}_1} & 0 \end{pmatrix} + \mathbf{u}_2 \begin{pmatrix} 0 & e^{i\mathbf{k}\cdot\mathbf{a}_2} \\ -e^{-i\mathbf{k}\cdot\mathbf{a}_2} & 0 \end{pmatrix} \Big]. \tag{A12}$$

When expressed in terms of $x$ and $y$ components as $\hat{\mathbf{J}}(\mathbf{k}) = \hat{x}\hat{J}_x(\mathbf{k}) + \hat{y}\hat{J}_y(\mathbf{k})$, we find (with $\mu = x, y$)

$$\hat{J}_\mu(\mathbf{k}) = \begin{pmatrix} 0 & j_\mu(\mathbf{k}) \\ j_\mu^*(\mathbf{k}) & 0 \end{pmatrix}, \tag{A13}$$

$$j_x(\mathbf{k}) = \sqrt{3}ta \sin\big(\frac{\sqrt{3}}{2}k_x a\big) e^{i\frac{3}{2}k_y a}, \tag{A14}$$

$$j_y(\mathbf{k}) = -ita \big[ \cos\big(\frac{\sqrt{3}}{2}k_x a\big) e^{i\frac{3}{2}k_y a} - 1 \big]. \tag{A15}$$

Note that although the physical electrical current is proportional to the electric charge $e$, we have suppressed this in our definition of the current density Eq. (A11). This can be easily reinstated in our final result for the conductivity.

## 2. Interaction Hamiltonian

The interaction Hamiltonian is:

$$H_{\text{int}} = \frac{e^2}{2} \sum_{\sigma\sigma'} \int d^3r\, d^3r' \frac{\psi_\sigma^\dagger(\mathbf{r})\, \psi_{\sigma'}^\dagger(\mathbf{r}')\, \psi_{\sigma'}(\mathbf{r}')\, \psi_\sigma(\mathbf{r})}{\epsilon |\mathbf{r} - \mathbf{r}'|}. \tag{A16}$$

Here, we have started with a full 3D Coulomb potential. Before proceeding, we note that the dimensionless parameter measuring the strength of the Coulomb interaction is the effective fine structure constant

$$\alpha = \frac{e^2}{\hbar v},\tag{A17}$$

with $v = \frac{3}{2}\frac{ta}{\hbar}$ the velocity parameter near the linear nodes of graphene. In the following, we aim to take into account the fact that electrons occupy orbitals on the honeycomb lattice. Therefore, we write the field-operator as:

$$\psi_\sigma(\mathbf{r}) = \sum_{\mathbf{R}_i} \sum_{\ell=1,2} c_{i\ell\sigma} \phi(\mathbf{r} - \mathbf{R}_i - \mathbf{v}_\ell),\tag{A18}$$

where the summation over $\ell$ refers to the two atoms in the unit cell, with $\mathbf{v}_\ell$ the vector pointing from the Bravais lattice site $\mathbf{R}_i$ to atom $\ell$. As in the main text, $\phi(\mathbf{r})$ is the 3D atomic orbital. When Eq. (A18) is inserted into Eq. (A16), we will have four lattice summations. However, we shall assume that products of factors like:

$$\phi^*(\mathbf{r} - \mathbf{R}_i - \mathbf{v}_\ell)\phi(\mathbf{r} - \mathbf{R}_{i'} - \mathbf{v}_{\ell'}),\tag{A19}$$

are only nonzero for $i = i'$ and $\ell = \ell'$, an approximation that amounts to assuming that the density of spin-$\sigma$ is approximated by:

$$\psi_\sigma^\dagger(\mathbf{r})\psi_\sigma(\mathbf{r}) = \sum_{\mathbf{R}_i} \sum_{\ell=1,2} c_{i\ell\sigma}^\dagger c_{i\ell\sigma} |\phi(\mathbf{r} - \mathbf{R}_i - \mathbf{v}_\ell)|^2.\tag{A20}$$

Using this approximation, and inserting the Fourier transform of the real-space Coulomb interaction $\frac{e^2}{\epsilon|\mathbf{r}-\mathbf{r}'|} = \int \frac{d^2q}{(2\pi)^2} \int \frac{dp_z}{2\pi} e^{i\mathbf{q}\cdot(\boldsymbol{\rho}-\boldsymbol{\rho}')} e^{ip_z(z-z')} \frac{4\pi e^2}{\epsilon(q^2+p_z^2)}$, we obtain:

$$H_{\text{int}} = \int d^3r \int d^3r' \int \frac{d^2q}{(2\pi)^2} \int \frac{dp_z}{2\pi} e^{i\mathbf{q}\cdot(\boldsymbol{\rho}-\boldsymbol{\rho}')} e^{ip_z(z-z')} \frac{4\pi e^2}{\epsilon(q^2+p_z^2)}$$
$$\times \sum_{\mathbf{R}_i,\ell} \sum_{\mathbf{R}_j,\ell'} c_{i\ell\sigma}^\dagger c_{j\ell'\sigma'}^\dagger c_{j\ell'\sigma'} c_{i\ell\sigma} |\phi(\mathbf{r} - \mathbf{R}_i - \mathbf{v}_\ell)|^2 |\phi(\mathbf{r}' - \mathbf{R}_j - \mathbf{v}_{\ell'})|^2,\tag{A21}$$

where we wrote $\mathbf{r} = (\boldsymbol{\rho}, z)$ and $\mathbf{r}' = (\boldsymbol{\rho}', z')$ with $\boldsymbol{\rho}$ a 2D coordinate in the plane of the honeycomb lattice and $z$ perpendicular to this plane. To evaluate the $\mathbf{r}$ and $\mathbf{r}'$ integrations, we first shift $\mathbf{r} \to \mathbf{r} + \mathbf{R}_i + v_\ell$ and similarly for $\mathbf{r}'$, an operation that will introduce phase factors of the form $e^{i\mathbf{q}\cdot(\mathbf{R}_i+\mathbf{v}_\ell)}$. The summations over $\mathbf{R}_i$ and $\mathbf{R}_j$ then implement lattice Fourier transforms on the operators $c_{i\ell\sigma}$, leading to:

$$H_{\text{int}} = \frac{1}{2} \int \frac{d^2q}{(2\pi)^2} V(\mathbf{q}) \sum_{\ell=1,2} \sum_{\ell'=1,2} e^{i\mathbf{q}\cdot(\mathbf{v}_\ell-\mathbf{v}_{\ell'})} \sum_{\mathbf{kk'}\sigma\sigma'} c_{\mathbf{k}+\mathbf{q}\ell\sigma}^\dagger c_{\mathbf{k}'-\mathbf{q}\ell'\sigma'}^\dagger c_{\mathbf{k}'\ell'\sigma'} c_{\mathbf{k}\ell\sigma}.\tag{A22}$$

$$= \frac{1}{2} \int \frac{d^2q}{(2\pi)^2} V(\mathbf{q}) \sum_{\mathbf{k}_1,\mathbf{k}_2} \hat{\psi}_k^\dagger(\mathbf{k}_2 - \mathbf{q})\hat{\psi}_i^\dagger(\mathbf{k}_1 + \mathbf{q})\hat{\psi}_j(\mathbf{k}_1)\hat{\psi}_\ell(\mathbf{k}_2) M_{ij}(\mathbf{q}) M_{k\ell}(-\mathbf{q}),\tag{A23}$$

where in the second line we switched to the notation of the preceding section, calling $c_{\mathbf{k}1\sigma} = a_{\mathbf{k}}$ and $c_{\mathbf{k}2\sigma} = b_{\mathbf{k}}$, and using the spinor Eq. (A8), with the indices $i$ and $j$ summing over the two spinor components. We also dropped the spin index, effectively studying spinless graphene (restoring spin-$\frac{1}{2}$ doubles the conductivity). Here, the matrix $M(\mathbf{q})$ is

$$M(\mathbf{q}) = \begin{pmatrix} 1 & 0 \\ 0 & e^{-i\mathbf{q}\cdot(\mathbf{v}_1-\mathbf{v}_2)} \end{pmatrix},\tag{A24}$$

and the Coulomb matrix element is

$$V(\mathbf{q}) = \int_{-\infty}^\infty \frac{dp_z}{2\pi} \frac{4\pi e^2 |\rho(\mathbf{q},p_z)|^2}{\epsilon(q^2 + p_z^2)}.\tag{A25}$$

In the following, we will use phenomenological forms for $V(\mathbf{q})$ aimed at investigating how the finite width of a Wannier function impacts the frequency-dependent conductivity. In the limiting case of pointlike atomic orbitals, we have $\rho(\mathbf{q}, p_z) \simeq 1$, implying for this case

$$V(\mathbf{q}) = \frac{2\pi e^2}{\epsilon q}.\tag{A26}$$

For the case of atomic orbitals that are pointlike in the $z$ direction but having a Gaussian shape in the 2D plane of graphene, $\phi(x, y) = \frac{1}{\lambda\sqrt{\pi}} \exp\left[-(x^2 + y^2)/2\lambda^2\right]$, we have

$$V(\mathbf{q}) = \frac{2\pi e^2}{\epsilon q} e^{-q^2\lambda^2/2}, \tag{A27}$$

clearly reducing to Eq. (A26) for $\lambda \to 0$.

### 3. Self energy

We expect that our tight-binding theory of graphene, that explicitly takes into account the nonzero spatial extent of the on-site Carbon orbitals, will not exhibit any ultraviolet divergences, despite the fact that conductivity contributions will involve integrations over all $\mathbf{q}$. In this section, we verify this directly for the case of the self energy.

Via standard perturbation theory, the self energy coming from Eq. (A23) is

$$\Sigma(\mathbf{p}) = -\int \frac{d^2q}{(2\pi)^2} V(\mathbf{q}) T \sum_\omega M(-\mathbf{q}) G(\mathbf{p} + \mathbf{q}, \omega) M(\mathbf{q}), \tag{A28}$$

After inserting the bare Green's function Eq. (A9) and evaluating the frequency sum, we find (defining $h(\mathbf{p}) = |h(\mathbf{p})|e^{i\phi(\mathbf{p})}$):

$$\begin{aligned}
\Sigma(\mathbf{p}) &= -\frac{1}{2} \int \frac{d^2q}{(2\pi)^2} V(\mathbf{q}) M(-\mathbf{q}) \begin{pmatrix} 0 & e^{i\phi(\mathbf{p}+\mathbf{q})} \\ e^{-i\phi(\mathbf{p}+\mathbf{q})} & 0 \end{pmatrix} M(\mathbf{q}), \\
&= -\frac{1}{2} \int \frac{d^2q}{(2\pi)^2} V(\mathbf{q}) \begin{pmatrix} 0 & e^{i[\phi(\mathbf{p}+\mathbf{q})-\mathbf{q}\cdot(\mathbf{v}_1-\mathbf{v}_2)]} \\ e^{-i[\phi(\mathbf{p}+\mathbf{q})-\mathbf{q}\cdot(\mathbf{v}_1-\mathbf{v}_2)]} & 0 \end{pmatrix}.
\end{aligned} \tag{A29}$$

Our next task is to show that the $\mathbf{q}$ integration in $\Sigma(\mathbf{p})$ is convergent. We consider the upper-right component

$$\Sigma_{12}(\mathbf{p}) = -\frac{1}{2} \int \frac{d^2q}{(2\pi)^2} V(\mathbf{q}) e^{i[\phi(\mathbf{p}+\mathbf{q})-\mathbf{q}\cdot(\mathbf{v}_1-\mathbf{v}_2)]}, \tag{A30}$$

$$= -\frac{1}{2} \int \frac{d^2q}{(2\pi)^2} V(\mathbf{q}) e^{-iq_y a} e^{i\phi(\mathbf{p}+\mathbf{q})}. \tag{A31}$$

where in the second line we made the choice $\mathbf{v}_1 = 0$, $\mathbf{v}_2 = \mathbf{u}_3 = -a\hat{y}$. Recall that the argument $\mathbf{p}$ of the self energy is defined within the BZ. We can extend this for $\mathbf{p}$ outside the BZ by defining $\Sigma_{12}(\mathbf{p})$ to be periodic in reciprocal lattice vectors $\mathbf{G}$, i.e., $\Sigma_{12}(\mathbf{p} + \mathbf{G}) = \Sigma_{12}(\mathbf{p})$. This further implies that we can define a self energy as a function of Bravais lattice position:

$$\Sigma_{12}(\mathbf{R}) = \int_{BZ} \frac{d^2p}{(2\pi)^2} \Sigma_{12}(\mathbf{p}) e^{-i\mathbf{p}\cdot\mathbf{R}}. \tag{A32}$$

Given $\Sigma_{12}(\mathbf{R})$, the momentum-dependent self energy is:

$$\Sigma_{12}(\mathbf{p}) = A \sum_{\mathbf{R}} e^{i\mathbf{p}\cdot\mathbf{R}} \Sigma_{12}(\mathbf{R}), \tag{A33}$$

where $A = \frac{3\sqrt{3}}{2}a^2$ is the unit-cell area. Now, $\Sigma_{12}(\mathbf{R})$ is:

$$\Sigma_{12}(\mathbf{R}) = -\frac{1}{2} \int_{BZ} \frac{d^2p}{(2\pi)^2} e^{-i\mathbf{p}\cdot\mathbf{R}} \int \frac{d^2q}{(2\pi)^2} V(\mathbf{q}) e^{-iq_y a} e^{i\phi(\mathbf{p}+\mathbf{q})} \tag{A34}$$

To proceed, we switch the order of integrations, doing the $\mathbf{p}$ integral first. We shift $\mathbf{p} \to \mathbf{p} - \mathbf{q}$ in the $\mathbf{p}$ integral, which is valid since the integrand is periodic in the BZ. This leads to:

$$\Sigma_{12}(\mathbf{R}) = -\frac{1}{2} \int \frac{d^2q}{(2\pi)^2} V(\mathbf{q}) e^{-iq_y a} e^{i\mathbf{q}\cdot\mathbf{R}} F(\mathbf{R}), \tag{A35}$$

$$F(\mathbf{R}) \equiv \int_{BZ} \frac{d^2p}{(2\pi)^2} e^{i[\phi(\mathbf{p})-\mathbf{p}\cdot\mathbf{R}]}. \tag{A36}$$

Now, the $\mathbf{q}$ integral in Eq. (A35) is independent of the function $F(\mathbf{R})$, and given by the Fourier transform of $V(\mathbf{q})$, evaluated at position $\mathbf{R} - a\hat{y}$. The result of this integration depends on the form of the Wannier function $\phi(\mathbf{r})$; our next task is to examine how the Wannier function affects the self energy by defining $V(\mathbf{q})$. As described above, pointlike Wannier functions on the honeycomb lattice imply Eq. (A26); for this case the integration in Eq. (A35) simply returns the real-space Coulomb interaction. Then, our result for $\Sigma_{12}(\mathbf{p})$ is:

$$\Sigma_{12}(\mathbf{p}) = -\frac{e^2}{2} A \sum_{\mathbf{R}} \mathrm{e}^{i\mathbf{p}\cdot\mathbf{R}} \frac{1}{|\mathbf{R} - a\hat{y}|} F(\mathbf{R}), \tag{A37}$$

a sum over Bravais lattice vectors. The advantage of this expression is that it does not feature a (numerically intensive) integration over all $\mathbf{q}$ like in Eq. (A31). Additionally, it is clear from this result that any divergences in $\Sigma_{12}$ cannot come from the Coulomb interaction (since the denominator of $\frac{1}{|\mathbf{R}-a\hat{y}|}$ never reaches zero) but only relies on the convergence of the sum at large $\mathbf{R}$. As we show below, one of the two main contributions to the conductivity relies on determining $\Sigma_{12}(\mathbf{p})$ for all momenta in the BZ by evaluating the sum in Eq. (A37). This result for $\Sigma_{12}(\mathbf{p})$ can also be generalized to the case of the Coulomb potential given in Eq. (A27), which incorporates a nonzero width to the on-site Wannier function, with the final result

$$\Sigma_{12}(\mathbf{p}) = -\frac{e^2}{2} A \sum_{\mathbf{R}} \mathrm{e}^{i\mathbf{p}\cdot\mathbf{R}} \sqrt{\frac{\pi}{2}} \frac{1}{\lambda} \mathrm{e}^{-|\mathbf{R}-a\hat{y}|^2/4\lambda^2} I_0\left(\frac{|\mathbf{R}-a\hat{y}|^2}{4\lambda^2}\right) F(\mathbf{R}), \tag{A38}$$

which reduces to Eq. (A37) in the limit $\lambda \ll a$. Note we have set the dielectric constant $\epsilon \to 1$ for simplicity; it may be easily reinstated by taking $e^2 \to e^2/\epsilon$.

### 4. Perturbative calculation of the conductivity

The frequency-dependent conductivity follows from the Kubo formula, which we now briefly review.[8] Using the Peierls substitution to couple an electromagnetic gauge field to electrons on the honeycomb lattice, we have:

$$\mathcal{H}_0(\mathbf{A}_i) = -t \sum_{\mathbf{R}_i} \sum_{n=1}^{3} \left( a^\dagger(\mathbf{R}_i) b(\mathbf{R}_i + \mathbf{u}_n) \mathrm{e}^{-i\mathbf{u}_n\cdot\mathbf{A}_i} + h.c. \right), \tag{A39}$$

with $A$ the unit cell area. Here, the electron charge has been set to unity. Taylor expanding to linear order gives

$$\mathcal{H}_0(\mathbf{A}_i) = \mathcal{H}_0 - A \sum_{\mathbf{R}_i} \mathbf{J}(\mathbf{R}_i) \cdot \mathbf{A}_i, \tag{A40}$$

$$\mathbf{J}(\mathbf{R}_i) = -i\frac{t}{A} \sum_{n=1}^{3} \left[ a^\dagger(\mathbf{R}_i) b(\mathbf{R}_i + \mathbf{u}_n) - b^\dagger(\mathbf{R}_i + \mathbf{u}_n) a(\mathbf{R}_i) \right]. \tag{A41}$$

Within time-dependent perturbation theory, the current-density at site $\mathbf{R}_i$ is:

$$\langle J_\mu(\mathbf{R}_i, t)\rangle = A \int_{-\infty}^{\infty} dt' \sum_{\mathbf{R}_j} \chi_{J,\mu,\nu}(\mathbf{R}_i, \mathbf{R}_j; t - t') A_\nu(\mathbf{R}_j), \tag{A42}$$

$$\chi_{J,\mu,\nu}(\mathbf{R}_i, \mathbf{R}_j; t - t') = i\Theta(t - t') \langle \left[ J_\mu(\mathbf{R}_i, t), J_\nu(\mathbf{R}_j, t') \right] \rangle, \tag{A43}$$

where in the second line we defined the retarded current-current correlator. Assuming the vector potential is uniform and has the time-dependence $\mathbf{A}(t) = \frac{\mathrm{e}^{-i\omega t}}{i\omega}\mathbf{E}$, with $\mathbf{E}$ the electric field, we find

$$\langle J_\mu \rangle = \sigma_{\mu\nu} E_\nu, \tag{A44}$$

$$\sigma_{\mu\nu}(\omega) = \frac{1}{\omega} \chi_{J,\mu,\nu}(\omega), \tag{A45}$$

where $\chi_{J,\mu,\nu}(\omega)$ is the spatial and temporal Fourier transform of Eq. (A43). As usual, this quantity can be obtained from the corresponding Matsubara function

$$\chi_{J,\mu,\nu}(i\Omega) = \frac{1}{NA} \int_0^\beta d\tau \mathrm{e}^{i\Omega\tau} \langle J_\mu(\tau) J_\nu(0) \rangle. \tag{A46}$$

In this formula $N$ is the number of Bravais lattice points, $A$ is the unit cell area, and $\beta = \frac{1}{k_B T}$ (although we always work in the zero-temperature limit). Here, $J_\mu(\tau)$ is given by Eq. (A11). Henceforth, we shall drop the subscript $\mu, \nu$ in the definition of $\chi_{J,\mu,\nu}(i\Omega)$, which we need for the case of $\mu = \nu$.

*a. Bare-bubble diagram*

To compute the conductivity, we need to evaluate $\chi_{J,\mu,\nu}(i\Omega)$ to leading order in perturbation theory. We start with the zeroth order result, which is the "bare-bubble" diagram, Fig. 2 (a) of the main text. Setting $\mu = \nu = y$ yields:

$$\chi_J^{(0)}(i\Omega) = -\int_{BZ} \frac{d^2k}{(2\pi)^2} \int \frac{d\omega}{2\pi} \text{Tr}\left[\hat{J}_y(\mathbf{k}, 0) G(\mathbf{k}, \omega) \hat{J}_y(\mathbf{k}) G(\mathbf{k}, \omega + \Omega)\right] \tag{A47}$$

$$= -\int_{BZ} \frac{d^2k}{(2\pi)^2} \frac{\left[h^*(\mathbf{k}) j_y(\mathbf{k}) - h(\mathbf{k}) j_y^*(\mathbf{k})\right]^2}{t \; |h(\mathbf{k})|(4|h(\mathbf{k})|^2 + \Omega^2/t^2)}. \tag{A48}$$

Here, $\hat{J}_\mu(\mathbf{k})$ is defined in Eq. (A13) and in the second line we evaluated the frequency integration and the trace. Next we rewrite the current-component $j_y(\mathbf{k})$, defined in Eq. (A15), as:

$$j_y(\mathbf{k}) = \frac{-ita}{2}\left[h(\mathbf{k}) - 3\right] \tag{A49}$$

$$j_y^*(\mathbf{k}) = \frac{ita}{2}\left[h^*(\mathbf{k}) - 3\right], \tag{A50}$$

and obtain for our retarded current correlator:

$$\chi_J^{(0)}(i\Omega) = \frac{ta^2}{4} \int_{BZ} \frac{d^2k}{(2\pi)^2} \frac{18|h(\mathbf{k})|^2 + 4|h(\mathbf{k})|^4 - 12|h(\mathbf{k})|^2 \left(h(\mathbf{k}) + h^*(\mathbf{k})\right) + 9\left(h(\mathbf{k})^2 + h^*(\mathbf{k})^2\right)}{|h(\mathbf{k})|(4|h(\mathbf{k})|^2 + \Omega^2/t^2)}. \tag{A51}$$

Upon analytically continuing $i\Omega \to \omega + i\delta$,

$$\frac{1}{4|h(\mathbf{k})|^2 + \Omega^2} \to \text{P.V.} \frac{1}{4|h(\mathbf{k})|^2 - \omega^2} + i\frac{\pi}{2\omega}\delta(\omega - 2|h(\mathbf{k})|), \tag{A52}$$

with P.V. denoting the principal value (and we assumed $\omega > 0$) and taking the imaginary part, we obtain the retarded correlator:

$$\chi_J^{(0)}(\omega) = \sum_{\mathbf{k}} \left(\frac{ta^2\pi}{32}\right) \left[18 + 4|h(\mathbf{k})|^2 + 18\frac{[\Re h(\mathbf{k})]^2 - [\Im h(\mathbf{k})]^2}{|h(\mathbf{k})|^2} - 24[\Re h(\mathbf{k})]\right] \delta\left(|h(\mathbf{k})| - \frac{\omega}{2t}\right) \tag{A53}$$

$$= \sum_{\mathbf{k}} \left(\frac{ta^2\pi}{32}\right) g\left(h(\mathbf{k})\right) \; \delta\left(|h(\mathbf{k})| - \frac{\omega}{2t}\right). \tag{A54}$$

In this expression, we have kept the dimensionful quantities $a$ and $t$, although henceforth we shall set them to unity and measure the frequency relative to $t$. Due to the delta function constraint, we can integrate the above expression analytically. Therefore we split up the function $g\left(h(\mathbf{k})\right)$ into two functions and define:

$$g_1(|h(\mathbf{k})|) = 18 + 4|h(\mathbf{k})|^2 \tag{A55}$$

$$g_2\left(h(\mathbf{k})\right) = 18\frac{[\Re h(\mathbf{k})]^2 - [\Im h(\mathbf{k})]^2}{|h(\mathbf{k})|^2} - 24[\Re h(\mathbf{k})]. \tag{A56}$$

Firstly, we evaluate the expression:

$$\chi_{J,1}^{(0)}(\omega) = \frac{\pi}{16} \sum_k g_1(|h(\mathbf{k})|)\delta(2|h(\mathbf{k})| - \omega). \tag{A57}$$

We introduce the density of state per unit cell as

$$\rho(E) = \int \frac{d^2k}{(2\pi)^2}\delta(E - |h(\mathbf{k})|)$$

$$= \int \frac{d^2k}{(2\pi)^2} \sum_{i=1}^4 \frac{1}{|\partial_{k_{x,i}}|h(k_{x,i}, k_y)||}\delta(k_x - k_{x,i}) \tag{A58}$$

with the $k_{x,i}$ being the solution to $E = |h(\mathbf{k})|$:

$$k_{x,1} = -\frac{2}{\sqrt{3}}\arccos\left[\frac{1}{4}(-2\cos\left(\frac{3k_y}{2}\right) - \sqrt{2}\sqrt{2E^2 - 1 + \cos(3k_y)})\right]$$

$$k_{x,2} = +\frac{2}{\sqrt{3}} \arccos\left[\frac{1}{4}(-2\cos\left(\frac{3k_y}{2}\right) - \sqrt{2}\sqrt{2E^2 - 1 + \cos(3k_y)})\right]$$

$$k_{x,3} = -\frac{2}{\sqrt{3}} \arccos\left[\frac{1}{4}(-2\cos\left(\frac{3k_y}{2}\right) + \sqrt{2}\sqrt{2E^2 - 1 + \cos(3k_y)})\right]$$

$$k_{x,4} = +\frac{2}{\sqrt{3}} \arccos\left[\frac{1}{4}(-2\cos\left(\frac{3k_y}{2}\right) + \sqrt{2}\sqrt{2E^2 - 1 + \cos(3k_y)})\right] \tag{A59}$$

describing curves that encircle the Dirac points at $\mathbf{k}_R = \frac{4\pi}{3a}(\frac{1}{2\sqrt{3}}\hat{x} + \frac{1}{2}\hat{y})$ and $\mathbf{k}_L = \frac{4\pi}{3a}(-\frac{1}{2\sqrt{3}}\hat{x} + \frac{1}{2}\hat{y})$ when the $y$ component is restricted to $k_- < k_y < k_+$ with

$$k_{\pm}(E) = \frac{2\pi}{3} \pm \frac{\arccos\left(1 - 2E^2\right)}{3} \ . \tag{A60}$$

We can calculate the density of states analytically and obtain:

$$\rho(E) = \frac{1}{(2\pi)^2} \frac{32 \ E \ \sqrt{1 - \frac{E}{3}} \ K\left[-\frac{16E}{(E-3)(1+E)^3}\right]}{3(3-E)(1+E)^{3/2}} \ , \tag{A61}$$

where $K[m]$ is the complete elliptic integral of the first kind. One part of the correlation function is thus given by:

$$\begin{aligned}
\chi^{(0)}_{J,1}(\omega) &= \frac{\pi}{32} \sum_{\mathbf{k}} g_1(|h(\mathbf{k})|)\delta(|h(\mathbf{k})| - \omega/2) \\
&= \frac{\pi}{32} \rho\left(\frac{\omega}{2}\right) g\left(\frac{\omega}{2}\right) \\
&= \frac{\pi}{32} \rho\left(\frac{\omega}{2}\right) (18 + \omega^2) \ . 
\end{aligned} \tag{A62}$$

In order to evaluate the expression:

$$\chi^{(0)}_{J,2}(\omega) = \frac{\pi}{16} \sum_{k} g_2(h(\mathbf{k}))\delta(2|h(\mathbf{k})| - \omega) \ , \tag{A63}$$

we expand the above formula near the node, $h(\mathbf{k}_R + \mathbf{k})$, and write the deviation from the node in polar coordinates $\mathbf{k} = (k, \theta)$,

$$|h(\mathbf{k}_R + \mathbf{k})| \simeq \frac{3}{128}k(64 - 7k^2 + 16k\cos 3\theta - k^2\cos 6\theta), \tag{A64}$$

valid to $\mathcal{O}(k^3)$. The approximate solution to $\omega = 2|h(\mathbf{k}_R + \mathbf{k})|$ is:

$$k_1(\theta, \omega) = \frac{1}{3}\omega - \frac{1}{36}\omega^2 \cos 3\theta + \frac{1}{1728}[7 + 8\cos^2 3\theta + \cos 6\theta]\omega^3 \ , \tag{A65}$$

that is valid to $\mathcal{O}(\omega^3)$. The factor $g_2(h(\mathbf{k}))$ is, to the same order,

$$\begin{aligned}
g_2(\mathbf{k}_R + \mathbf{k}) \simeq \frac{9}{32}[&k^3(-\cos 11\theta) + 3\left(5k^2 - 16\right)k\cos\theta + \left(64 - 20k^2\right)\cos 2\theta \\
&+ 2k\left(2\left(k^2 - 16\right)\cos 3\theta - 8(\cos 5\theta + 3k) + k(8\cos 4\theta + 8\cos 6\theta + 3k\cos 5\theta + 2\cos 8\theta(1 - 2k\cos\theta))\right)] \ . 
\end{aligned} \tag{A66}$$

From the delta function, we'll also need

$$\frac{d}{dk}|h(\mathbf{k}_R + \mathbf{k})| = \frac{3}{64}\left(64 + 32k\cos 3\theta - 21k^2 - 3k^2\cos 6\theta\right) \ . \tag{A67}$$

Then, assuming the same contribution comes from each node (which we have verified), we'll have:

$$\chi^{(0)}_{J,2}(\omega) = \frac{\pi}{8} \int\limits_0^{2\pi} d\theta \int\limits_0^\infty dk \ k \ g_2(\mathbf{k})\delta(2|h(\mathbf{k})| - \omega) \tag{A68}$$

$$= \frac{\pi}{8} \int\limits_0^{2\pi} d\theta \ k_1(\theta, \omega)\frac{1}{|\frac{d}{dk_1}2|h(\mathbf{k}_R + \mathbf{k}_1)||} g_2(\mathbf{k}_R + \mathbf{k}_1) \ , \tag{A69}$$

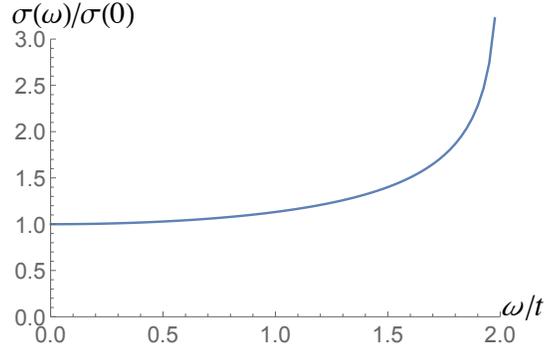

FIG. 2. Bare-bubble conductivity $\sigma(\omega)$, normalized to its value at zero frequency.

where we evaluated the radial $k$ integral. To evaluate the integral, we simply insert $k_1(\theta, \omega)$ into the factors Eq. (A67) and Eq. (A68), insert them into the integrand and Taylor expand order by order in $\omega$ before evaluating the angle integrations. We obtain:

$$\chi_{J,2}^{(0)}(\omega) = -\frac{\omega}{8}\frac{\omega^2}{36}. \tag{A70}$$

Upon inserting the combined result into Eq. (A45), we find the frequency-dependent conductivity plotted in Fig. 2 and given by the formula:

$$\sigma(\omega) = \frac{\pi}{32\omega}\rho\left(\frac{\omega}{2}\right)(18 + \omega^2) - \frac{1}{8}\frac{\omega^2}{36} \tag{A71}$$

$$\approx \sigma_0\left(1 + \frac{1}{9}\omega + \mathcal{O}(\omega^3)\right), \tag{A72}$$

with $\sigma_0$ the zero-frequency limit (reinserting correct factors of $e^2$ and $\hbar$, previously set to unity).

$$\sigma_0 = \frac{1}{8}\frac{e^2}{\hbar}. \tag{A73}$$

In comparing to the known result for the conductivity of $N$ species of Dirac fermions, $\sigma_0 = \frac{N}{16}\frac{e^2}{\hbar}$, recall that here we have $N = 2$, since we are considering the spinless case (but have summed over two nodes).

### b. Interaction corrections to the conductivity

The leading order interaction corrections to the conductivity, that are linear order in the effective fine structure constant $\alpha$, can be expressed in terms of self-energy (diagrams $b$ and $c$) and vertex type (diagam $d$) Feynman diagrams, as depicted in Fig. 2 ($a$) and ($b$) of the main text. The self-energy contribution is:

$$\chi_J^{(1)}(i\Omega)\Big|_{bc} = -2\sum_{\mathbf{k}}\int\frac{d\omega}{2\pi}G(\mathbf{k}, \omega - \Omega)J_\mu(\mathbf{k})G(\mathbf{k}, \omega)\Sigma(\mathbf{k})G(\mathbf{k}, \omega)J_\mu(\mathbf{k}), \tag{A74}$$

with the overall 2 coming from there being two such diagrams. The latter is given by:

$$\chi_J^{(1)}(i\Omega)\Big|_d = \sum_{\mathbf{k}}\int\frac{d^2q}{(2\pi)^2}V(\mathbf{q})\mathrm{Tr}\left[I_\mu(\mathbf{k}, \Omega)M(-\mathbf{q})I_\mu(\mathbf{k} + \mathbf{q}, -\Omega)M(\mathbf{q})\right], \tag{A75}$$

with

$$I_\mu(\mathbf{k}, \Omega) = \int\frac{d\omega}{2\pi}G(\mathbf{k}, \omega)J_\mu(\mathbf{p})G(\mathbf{k}, \omega - \Omega). \tag{A76}$$

We emphasize that momentum summations are always over the Brillouin zone and $\mathbf{q}$ integrations are always over the entire 2D momentum space, an issue that was neglected in Ref. 9 (see Eq.(15) of this work). As in our calculation of

the self energy, a crucial simplification of Eqs. (A74) and (A75) will involve writing these in a way that allows us to analytically evaluate the $\mathbf{q}$ integration.

Before analyzing these results in the subsequent sections, we first recall the result for the $bc$ and $d$ contributions within the nodal approximation. In this approximation, regularizing the integals by imposing a large momentum cutoff $\Lambda$ on the Coulomb potential (a procedure which obeys the Ward identity[4]), we obtain:

$$\text{Im}\chi_J^{(1)}\Big|_{bc}(\omega) = -\frac{1}{2}\alpha\omega\sigma_0 \ln\frac{8\Lambda v}{\omega}, \tag{A77}$$

with $v = 3ta/2\hbar$, and

$$\text{Im}\chi_J^{(1)}\Big|_d(\omega) = \frac{1}{2}\alpha\omega\sigma_0\Big[\ln\frac{8\Lambda v}{\omega} + \frac{19-6\pi}{6}\Big], \tag{A78}$$

yielding the sum

$$\text{Im}\chi_J^{(1)}\Big|_{bc}(\omega) + \text{Im}\chi_J^{(1)}\Big|_d(\omega) = \sigma_0\alpha\frac{19-6\pi}{12}. \tag{A79}$$

These formulas for the $bc$ and $d$ contributions are of course approximately valid within the tight-binding theory, with the replacement of the UV cutoff $\Lambda \to 1/a$, so that we expect each contribution to go as $\sim \omega\ln\omega$. This creates numerical difficulties, as each term is large, requiring a cancellation to return the value that is consistent with the nodal result $\propto \frac{19-6\pi}{12} \simeq 0.0125$. However, as shown in the main text, our numerical calculations are indeed consistent with this value.

### c. Diagrams b and c

Starting with Eq. (A74), we first evaluate the frequency integration and the trace. Then we find (summing over the $xx$ and $yy$ components and dividing by 2):

$$\chi_J^{(1)}(i\Omega)\Big|_{bc} = -\int_{\text{BZ}}\frac{d^2p}{(2\pi)^2}\frac{1}{4|h(\mathbf{p})|^3}\Big[\frac{D_1(\mathbf{p})}{4|h(\mathbf{p})|^2+\Omega^2} + D_2(\mathbf{p})\frac{4|h(\mathbf{p})|^2-\Omega^2}{(4|h(\mathbf{p})|^2+\Omega^2)^2}\Big], \tag{A80}$$

where the functions $D_1(\mathbf{p})$ and $D_2(\mathbf{p})$ are given by:

$$D_1(\mathbf{p}) = 2\big(h^*\Sigma_{12} - h\Sigma_{21}\big)\big[(h^*)^2(j_x^2+j_y^2) - h^2((j_x^*)^2+(j_y^*)^2)\big], \tag{A81}$$

$$D_2(\mathbf{p}) = \big(h^*\Sigma_{12} + h\Sigma_{21}\big)\Big[(h^*)^2(j_x^2+j_y^2) + h^2((j_x^*)^2+(j_y^*)^2) - 2(j_xj_x^* + j_yj_y^*)|h|^2\Big] \tag{A82}$$

Although this expression is complicated, all that is left is to analytically continue $i\Omega \to \omega+i\delta$ and take the imaginary part. The analytical continuation can be performed using Eq. (A52) for the term proportional to $D_1(\mathbf{p})$ and

$$\frac{4|h|^2-\Omega^2}{(4|h|^2+\Omega^2)^2} = \frac{d}{d\Omega}\frac{\Omega}{4|h|^2+\Omega^2} \to \frac{d}{d\omega}\frac{\omega}{4|h|}\Big[\frac{1}{2|h|+\omega+i\delta} + \frac{1}{2|h|-\omega-i\delta}\Big], \tag{A83}$$

for the term proportional to $D_2(\mathbf{p})$. After taking the imaginary part (and assuming $\omega > 0$), we have

$$\text{Im}\chi_J^{(1)}\Big|_{bc}(\omega) = -\pi\int_{\text{BZ}}\frac{d^2p}{(2\pi)^2}\frac{1}{16|h(\mathbf{p})|^4}D_1(\mathbf{p})\delta(\omega-2|h(\mathbf{p})|) - \pi\frac{d}{d\omega}\int_{\text{BZ}}\frac{d^2p}{(2\pi)^2}\frac{\omega}{16|h(\mathbf{p})|^4}D_2(\mathbf{p})\delta(\omega-2|h(\mathbf{p})|). \tag{A84}$$

To evaluate this, then, we determine $\Sigma_{12}(\mathbf{k})$ for $\mathbf{k}$ within the BZ by evaluating the summation over $\mathbf{R}$ for a large set of Bravais lattice vectors. With the delta function constraint, all that remains is a numerical integration over $p$ along the curves $\omega = 2|h(\mathbf{p})|$ (which go around the Dirac nodes).

### d. Diagram d

Next, we turn to Eq. (A75). Our first task is to evaluate Eq. (A76). We find:

$$I_\mu(\mathbf{p},\Omega) = \frac{1}{|h(\mathbf{p})|(4|h(\mathbf{p})|^2+\Omega^2)}V_\mu(\mathbf{p},\Omega), \tag{A85}$$

$$V_\mu(\mathbf{p}, \Omega) \equiv \begin{pmatrix} \frac{1}{2}i\Omega\big[h(\mathbf{p})j_\mu^*(\mathbf{p}) - h^*(\mathbf{p})j_\mu(\mathbf{p})\big] & h(\mathbf{p})^2 j_\mu^*(\mathbf{p}) - |h(\mathbf{p})|^2 j_\mu(\mathbf{p}) \\ h^*(\mathbf{p})^2 j_\mu(\mathbf{p}) - |h(\mathbf{p})|^2 j_\mu^*(\mathbf{p}) & \frac{1}{2}i\Omega\big[h^*(\mathbf{p})j_\mu(\mathbf{p}) - h(\mathbf{p})j_\mu^*(\mathbf{p})\big] \end{pmatrix}. \tag{A86}$$

Now, we have

$$\chi_J^{(1)}\Big|_d = \int \frac{d^2q}{(2\pi)^2} V(\mathbf{q}) \int_{BZ} \frac{d^2p}{(2\pi)^2} \frac{1}{|h(\mathbf{p})|(4|h(\mathbf{p})|^2 + \Omega^2)} \frac{1}{|h(\mathbf{p}+\mathbf{q})|(4|h(\mathbf{p}+\mathbf{q})|^2 + \Omega^2)}$$
$$\times \mathrm{Tr}\big[V_\mu(\mathbf{p}, \Omega) M(-\mathbf{q}) V_\mu(\mathbf{p}+\mathbf{q}, -\Omega) M(\mathbf{q})\big], \tag{A87}$$

which we now proceed to simplify. Recall that, in the $bc$ diagram, we expressed the self energy as a summation over Bravais lattice vectors, so that the $\mathbf{q}$ integration did not need to be performed numerically (i.e., it was performed analytically to yield the real-space Coulomb interaction). In the present case of Eq. (A87), we can perform a similar trick by first writing the integral as

$$\chi_J^{(1)}\Big|_d = = \int_{BZ} \frac{d^2p}{(2\pi)^2} \frac{1}{|h(\mathbf{p})|(4|h(\mathbf{p})|^2 + \Omega^2)} \mathrm{Tr}\big[V_\mu(\mathbf{p}, \Omega) Q_\mu(\mathbf{p}, -\Omega)\big], \tag{A88}$$

$$Q_\mu(\mathbf{p}, -\Omega) \equiv \int \frac{d^2q}{(2\pi)^2} V(\mathbf{q}) \frac{M(-\mathbf{q}) V_\mu(\mathbf{p}+\mathbf{q}, -\Omega) M(\mathbf{q})}{|h(\mathbf{p}+\mathbf{q})|(4|h(\mathbf{p}+\mathbf{q})|^2 + \Omega^2)}, \tag{A89}$$

Much like the self-energy, we can express $Q_\mu(\mathbf{p}, -\Omega)$ as a sum over Bravais lattice vectors of a summand for which the $\mathbf{q}$ integration may be performed analytically. The result is:

$$Q_\mu(\mathbf{p}, -\Omega) = A \sum_{\mathbf{R}} e^{i\mathbf{p}\cdot\mathbf{R}} \int_{BZ} \frac{d^2p'}{(2\pi)^2} e^{-i\mathbf{p}'\cdot\mathbf{R}} \begin{pmatrix} V_{\mu,11}(\mathbf{p}', -\Omega)\frac{e^2}{|\mathbf{R}|} & V_{\mu,12}(\mathbf{p}', -\Omega)\frac{e^2}{|\mathbf{R}-a\hat{y}|} \\ V_{\mu,21}(\mathbf{p}', -\Omega)\frac{e^2}{|\mathbf{R}+a\hat{y}|} & V_{\mu,22}(\mathbf{p}', -\Omega)\frac{e^2}{|\mathbf{R}|} \end{pmatrix} \frac{1}{|h(\mathbf{p}')|(4|h(\mathbf{p}')|^2 + \Omega^2)}, \tag{A90}$$

which now involves a summation over Bravais lattice vectors and an integration over the BZ. Inserting this into, Eq. (A87), evaluating the trace, and simplifying, we find:

$$\chi_J(i\Omega)\Big|_d = e^2 A \sum_{\mathbf{R}} \int_{BZ} \frac{d^2p}{(2\pi)^2} \int_{BZ} \frac{d^2p'}{(2\pi)^2} e^{i(\mathbf{p}-\mathbf{p}')\cdot\mathbf{R}} \frac{1}{|h(\mathbf{p})|(4|h(\mathbf{p})|^2 + \Omega^2)} \frac{1}{|h(\mathbf{p}')|(4|h(\mathbf{p}')|^2 + \Omega^2)} \tag{A91}$$
$$\times \left[ \frac{V_{\mu,11}(\mathbf{p}, \Omega) V_{\mu,11}(\mathbf{p}', -\Omega) + V_{\mu,22}(\mathbf{p}, \Omega) V_{\mu,22}(\mathbf{p}', -\Omega)}{|\mathbf{R}|} + \frac{V_{\mu,21}(\mathbf{p}, \Omega) V_{\mu,12}(\mathbf{p}', -\Omega)}{|\mathbf{R} - a\hat{y}|} + \frac{V_{\mu,12}(\mathbf{p}, \Omega) V_{\mu,21}(\mathbf{p}', -\Omega)}{|\mathbf{R} + a\hat{y}|} \right].$$

The next step is to analytically continue, using Eq. (A52). For the imaginary part, we'll clearly have two terms, one with $\delta(\omega - 2|h(\mathbf{p})|)$ and one with $\delta(\omega - 2|h(\mathbf{p}')|)$. However, since the integrand is symmetric under exchanging $\mathbf{p}$ and $\mathbf{p}'$ and also $\mathbf{R} \to -\mathbf{R}$, these two terms are identical. We finally obtain (summing over the $xx$ and $yy$ components and dividing by 2, and using Eq. (A86)) :

$$\chi_J(\omega)\Big|_d = -e^2 A \sum_{\mu=x,y} \sum_{\mathbf{R}} \int_{BZ} \frac{d^2p}{(2\pi)^2} \int_{BZ} \frac{d^2p'}{(2\pi)^2} e^{i(\mathbf{p}-\mathbf{p}')\cdot\mathbf{R}} \big[h(\mathbf{p})j_\mu^*(\mathbf{p}) - h^*(\mathbf{p})j_\mu(\mathbf{p})\big]\big[h(\mathbf{p}')j_\mu^*(\mathbf{p}') - h^*(\mathbf{p}')j_\mu(\mathbf{p}')\big]$$
$$\times \left[ \frac{\omega^2}{2|\mathbf{R}|} + \frac{h^*(\mathbf{p})h(\mathbf{p}')}{|\mathbf{R} - a\hat{y}|} + \frac{h(\mathbf{p})h^*(\mathbf{p}')}{|\mathbf{R} + a\hat{y}|} \right] \frac{\pi}{2\omega} \delta(\omega - 2|h(\mathbf{p})|) \frac{1}{|h(\mathbf{p})||h(\mathbf{p}')|} \mathrm{P.V.} \frac{1}{4|h(\mathbf{p}')|^2 - \omega^2}, \tag{A92}$$

the result for the retarded correlator. To evaluate this diagram, we must numerically evaluate the momentum integrations over the Brillouin zone and the summation over BL vectors $\mathbf{R}$. This result for pointlike Wannier functions can again be generalized to the case of the Coulomb potential given in Eq. (A27). Upon substituting the expression $\frac{1}{|\mathbf{R}-a\hat{y}|}$ by $\sqrt{\frac{\pi}{2}} \frac{1}{\lambda} e^{-|\mathbf{R}-a\hat{y}|^2/4\lambda^2} I_0\left(\frac{|\mathbf{R}-a\hat{y}|^2}{4\lambda^2}\right)$ and analogously for $\frac{1}{|\mathbf{R}|}$ and $\frac{1}{|\mathbf{R}+a\hat{y}|}$, the nonzero width of the on-site Wannier function is taken into account.

## Appendix B: Dirac theory of graphene

Within the Dirac theory of graphene, both contributions to the interaction correction to the conductivity involve divergent integrals. In this section, our main goal is to reconcile the results of dimensional regularization[6] with the tight-binding results. We first describe (following Teber and Kotikov (TK)[7]) the dimensional regularization (DR) of the Dirac theory of graphene, based on the modified minimal subtraction $\overline{MS}$ scheme, and the corresponding

continuum renormalization group (RG). As found by TK, although the interaction correction diagrams $b, c$ and $d$ yield a different result $\mathcal{C}'$, within DR the bare bubble contribution is also renormalized, giving an additional correction that finally yields the same result for the interaction correction.

Subsequently, we compute the conductivity of graphene in $d = 2 - \epsilon$ dimensions but including a UV cutoff $\Lambda$, showing that the limits of $\epsilon \to 0$ and $\Lambda \to \infty$ do not commute, a phenomenon that we refer to as an ultraviolet (UV) quirk. We do this using the relation between the conductivity and density-density correlator used in the original work of Mishchenko.[2] This allows us to precisely isolate the origin of the UV quirk as coming from the $bc$ diagrams, and we show that, if one has a UV cutoff in $d$ dimensions, the result $\mathcal{C}$ is obtained when the limit $d \to 2$ is taken. However, if one works in $\epsilon$ dimensions but with $\Lambda \to \infty$, then the $bc$ diagram yields a different result. In this case, the theory must be regularized using the $\overline{MS}$ scheme, which, again, finally yields the result proportional to $\mathcal{C}$.

### 1. Continuum RG/ Minimal subtraction scheme

The results of this section largely follow the work of TK,[7] in particular highlighting the differences relative to the momentum shell RG. In this section, we use the notation $D$ to refer to the spatial dimensionality (in contrast to $d$ which is used elsewhere). When using dimensional regularization, we do not introduce an UV-cutoff $\Lambda$ to our system. The system is regularized by the dimensionless parameter $\frac{1}{\epsilon}$. Thus applying the dimensional regularization scheme to our Feynman diagrams, namely the vertex-diagram and the self-energy diagram, we obtain for the correction coefficient $\mathcal{C}' = \frac{22-6\pi}{12}$. But the whole theory is divergent and needs to be regularized by the continuum renormalization group. This is done by introducing counter terms which remove these divergences. After this regularization procedure, we obtain as value for the correction coefficient $\mathcal{C} = \frac{19-6\pi}{12}$.

Our starting point is the action of graphene:

$$S = \int d\tau \int d^{D_e}x \ \psi_0^\dagger \left(\partial_\tau + \mathrm{i}e_0 A_0^0 + v_0(-\mathrm{i}\boldsymbol{\nabla}\boldsymbol{\sigma})\right)\psi_0 + \int d\tau \int d^{D_\gamma}x \left(\partial_x A_0^0\right)^2 \ , \tag{B1}$$

where the subscript $X_0$ denotes the bare quantities. Here, $D_e$ is the spatial dimensionality of the electron degrees of freedom (henceforth we call $D_e \to D = 2 - \epsilon$) and $D_\gamma = 3$ is the dimensionality of the gauge fields $A_0$ mediating the Coulomb potential. The corresponding Lagrangian of the bare physical quantities is given by:

$$\mathcal{L}_0 = \psi_0^\dagger \partial_\tau \psi_0 + \mathrm{i}e_0 \psi_0^\dagger A_0^0 \psi_0 + v_0 \psi_0^\dagger(-\mathrm{i}\boldsymbol{\nabla}\boldsymbol{\sigma})\psi_0 + \left(\partial_x A_0^0\right)^2 \ . \tag{B2}$$

The Coulomb- interaction of graphene in $D = 2 - \epsilon$ dimensions is given by:

$$V(\mathbf{r}) = \frac{e^2}{r} \left(\frac{r}{r_0}\right)^\epsilon \ , \tag{B3}$$

where we introduced the length scale $r_0$ in such a way, that the Coulomb potential has the correct units in $D = 2 - \epsilon$. Upon Fourier transforming this potential in $D = 2 - \epsilon$ dimensions, we obtain:

$$V(\mathbf{q}) = \frac{2\pi e^2}{q} \frac{r_0^{-\epsilon}\pi^{-\epsilon/2}\Gamma\left(\frac{1}{2}\right)}{\Gamma\left(\frac{1-\epsilon}{2}\right)} \ . \tag{B4}$$

This theory contains divergences, which we can see when we calculate the self-energy in $D = 2 - \epsilon$ dimensions:

$$\begin{aligned}
\Sigma(p) &= \Phi(p) \ v_0 \mathbf{p}\boldsymbol{\sigma} \\
&= \frac{e_0^2}{v_0} \frac{\alpha 2^{2\epsilon-3}\Gamma\left(\frac{\epsilon}{2}\right)}{\Gamma\left(1-\frac{\epsilon}{2}\right)}(r_0 p)^{-\epsilon} \ v_0 \mathbf{p}\boldsymbol{\sigma} \\
&= \alpha_0 \ \frac{\alpha 2^{2\epsilon-3}\Gamma\left(\frac{\epsilon}{2}\right)}{\Gamma\left(1-\frac{\epsilon}{2}\right)}(r_0 p)^{-\epsilon} \ v_0 \mathbf{p}\boldsymbol{\sigma} \ .
\end{aligned} \tag{B5}$$

If we expand the function $\Phi(p)$ for small $\epsilon$ to order $\mathcal{O}(\epsilon^0)$ we obtain:

$$\Phi(p) \approx \frac{\alpha_0}{4\epsilon} + \frac{\alpha_0}{4}\left(\log(4) - \log(pr_0) - \gamma\right) \ . \tag{B6}$$

Here we explicitly see that the theory has divergences. In order to make the theory finite, we introduce the renormalized, physical fields, which we denote with a subscript (e.g., $X_R$ for observable $X$). The fields are renormalized the following way:

$$\psi_0 = \sqrt{Z_\psi} \ \psi_R, \tag{B7}$$

$$A_0 = \sqrt{Z_A} \, A_R, \tag{B8}$$

$$v_0 = Z_v \, v_R \,. \tag{B9}$$

The renormalized Lagrangian has now the form:

$$\mathcal{L}_R = Z_\psi \, \psi_R^\dagger \partial_\tau \psi_R + \mathrm{i}e_0 \sqrt{Z_A} Z_\psi \, \psi_R^\dagger A_R^0 \psi_R + Z_v Z_\psi \, v_R \psi_R^\dagger (-\mathrm{i}\boldsymbol{\nabla}\boldsymbol{\sigma}) \psi_R + Z_A \, \left(\partial_x A_R^0\right)^2 \,, \tag{B10}$$

where we introduce the conventional definition:

$$Z_1 = \frac{e_0}{e_R} \, Z_\psi \sqrt{Z_A} \;\Leftrightarrow\; e_0 = e_R \frac{Z_1}{Z_\psi \sqrt{Z_A}} = e_R \, Z_e \,. \tag{B11}$$

The renormalized Lagrangian is now:

$$\mathcal{L} = Z_\psi \, \psi_R^\dagger \partial_\tau \psi_R + Z_e \, \mathrm{i}e_R \psi_R^\dagger A_R^0 \psi_R + Z_v Z_\psi \, v_R \psi_R^\dagger (-\mathrm{i}\boldsymbol{\nabla}\boldsymbol{\sigma}) \psi_R + Z_A \, \left(\partial_x A_R^0\right)^2 \,, \tag{B12}$$

Next we introduce the counter terms which make our theory finite. They are defined the following way:

$$Z_\psi = 1 + \delta_\psi, \tag{B13}$$
$$Z_A = 1 + \delta_A, \tag{B14}$$
$$Z_v = 1 + \delta_v, \tag{B15}$$
$$Z_e = 1 + \delta_e \,. \tag{B16}$$

The Lagrangian has now the form:

$$\begin{aligned}
\mathcal{L}_R =\; & \psi_R^\dagger \partial_\tau \psi_R + \mathrm{i}e_R \psi_R^\dagger A_R^0 \psi_R + v_R \psi_R^\dagger (-\mathrm{i}\boldsymbol{\nabla}\boldsymbol{\sigma}) \psi_R + \left(\partial_x A_R^0\right)^2 \\
& + \delta_\psi \, \left(\psi_R^\dagger \partial_\tau \psi_R + v_R \psi_R^\dagger (-\mathrm{i}\boldsymbol{\nabla}\boldsymbol{\sigma}) \psi_R\right) + \delta_e \, \mathrm{i}e_R \psi_R^\dagger A_R^0 \psi_R + \delta_v \, v_R \psi_R^\dagger (-\mathrm{i}\boldsymbol{\nabla}\boldsymbol{\sigma}) \psi_R + \delta_A \, \left(\partial_x A_R^0\right)^2 \,.
\end{aligned} \tag{B17}$$

The counterterms are chosen in such a way that they cancel the divergences. In the case of the self-energy we have:

$$\Sigma(p) \sim \Phi_R(p) \, v_R \mathbf{p}\boldsymbol{\sigma} + \delta_v \, v_R \mathbf{p}\boldsymbol{\sigma} + \delta_\psi \, (\mathrm{i}\Omega + v_R \mathbf{p}\boldsymbol{\sigma}) \,. \tag{B18}$$

We have seen in equation (B6), that the divergence of the self-energy is independent of the frequency. We must thus choose the following condition for $\delta_\psi$:

$$\delta_\psi = 0 \;\rightarrow\; Z_\psi = 1 \,. \tag{B19}$$

Thus, examining Eq. (B18), the velocity counterterm must cancel the divergence of the self-energy. We have:

$$\delta_v = -\frac{\alpha_R}{4} \frac{1}{\epsilon} \;\rightarrow\; Z_v = \left(1 - \frac{\alpha_R}{4\epsilon}\right) \,. \tag{B20}$$

The electrical charge stays unrenormalized in graphene, i.e. $Z_e = 1$. Furthermore, the electrical charge has the dimensionality

$$[e] = \epsilon \,. \tag{B21}$$

This can be derived by taking a closer look at the action of graphene, which has to be a dimensionless quantity. This leads to the fact, that the ferminonic fields have the following dimensionality:

$$[\psi_0] = \frac{D_e}{2} \,, \tag{B22}$$

while the bosonic fields of the photons have the dimension:

$$[A_0^\mu] = 1 - \epsilon \,. \tag{B23}$$

We can now deduce from these conditions that the dimension of the electrical charge $e_0$ is given by:

$$[e_0] = \epsilon \,. \tag{B24}$$

In the modified minimal substraction scheme ($\overline{MS}$), we introduce a physical scale $\mu$ in such a way that the physical observable becomes dimensionless and the divergence is removed.[5] We have in the case of the self-energy the following:

$$\Phi(p) = \frac{\alpha_R}{4\epsilon} + \frac{\alpha_R}{4}\left(\log(4) - \log(pr_0) - \gamma\right) \tag{B25}$$

$$\rightarrow \frac{\alpha_R}{4}\left(-\log\left(\frac{pr_0}{\tilde{\mu}/\omega}\right)\right), \tag{B26}$$

with

$$\tilde{\mu} = 4e^{-\gamma}\mu. \tag{B27}$$

In other words, we substitute the divergence by a logartithm:

$$\frac{1}{\epsilon} \rightarrow \log\left(\mu/\omega\right). \tag{B28}$$

Next, we study the electrical charge more closely. We have seen in Eq. (B24) that the bare charge has dimensionality $[e_0] = \epsilon$. However, in order to have the electrical charge as a dimensionless quantity, we introduce again the parameter $\mu$:

$$\frac{e_0^2}{\mu^{2\epsilon}} = e^2(\mu)Z_e \iff e_0^2 = \frac{e^2(\mu)}{(4)^\epsilon}e^{2\gamma\epsilon}\tilde{\mu}^{-2\epsilon} \iff e^2(\mu) = e_0^2(4)^\epsilon e^{-2\gamma\epsilon}\tilde{\mu}^{-2\epsilon}. \tag{B29}$$

Next we recall that the velocity is renormalized by:

$$v_0 = \left[1 - \frac{1}{4\epsilon}\alpha(\mu)\right]v(\mu), \tag{B30}$$

which can be rewritten as:

$$v(\mu) = \frac{4\epsilon v_0}{4\epsilon - \alpha(\mu)}. \tag{B31}$$

Now we can define our coupling constant as:

$$\alpha(\mu) = \frac{e^2(\mu)}{v(\mu)}. \tag{B32}$$

Combining the above equation with equation (B29) and equation (B31), we obtain for the coupling constant the following expression:

$$\alpha(\mu) = \frac{\alpha_0 4^\epsilon e^{-2\gamma\epsilon}\tilde{\mu}^{-2\epsilon}}{1 + \frac{\alpha_0}{4}\frac{1}{\epsilon}4^\epsilon e^{-2\gamma\epsilon}\tilde{\mu}^{-2\epsilon}}. \tag{B33}$$

Next we replace again the divergence by the logarithm, using $\frac{1}{\epsilon} \rightarrow \log\left(\mu/\omega\right)$, and than take the limit $\epsilon \rightarrow 0$. This yields:

$$\alpha(\omega) = \frac{\alpha_0}{1 + \frac{\alpha_0}{4}\log\left(\frac{\mu}{\omega}\right)}. \tag{B34}$$

The velocity is treated analogous. It holds:

$$v_0 = v(\mu) - \frac{1}{4\epsilon}e^2(\mu) \implies v(\mu) = v_0 + \frac{1}{4\epsilon}e_0^2 4^\epsilon e^{-2\gamma\epsilon}\tilde{\mu}^{-2\epsilon}. \tag{B35}$$

After replacing the divergence and taking the limit $\epsilon \rightarrow 0$, we obtain:

$$v(\omega) = v_0 + \frac{e_0^2}{4}\log\left(\frac{\mu}{\omega}\right). \tag{B36}$$

At last we can study the non-interacting optical conductivity $\sigma_0$, which requires computing the bare-bubble diagram in spatial dimension $d$. We find:

$$\sigma_{0,0}(\omega) = e_0^2 N A_\epsilon \left|\frac{v_0}{\omega}\right|^\epsilon, \tag{B37}$$

where

$$A_\epsilon = 4^{-2+\epsilon} \pi^{\epsilon/2} \frac{(1-\epsilon)}{\Gamma\left(1-\frac{\epsilon}{2}\right)} \,. \tag{B38}$$

In order to obtain a theory which does not have divergences we have to replace all bare quantities (such as the charge and velocity parameters) by their renormalized values. This prodedure gives us the following expression:

$$\sigma_{0,R}(\omega) = \frac{e^2(\mu)}{4^\epsilon} e^{2\gamma\epsilon} \tilde{\mu}^{2\epsilon} N A_\epsilon \left| \frac{\left(1 - \frac{\alpha(\mu)}{4\epsilon}\right) v_R}{\omega} \right|^\epsilon \,. \tag{B39}$$

According to equation (B34), we can choose $\alpha(\mu)$ in such a way that it is a small quantity. Using this we can approximate[7]

$$\left(1 - \frac{\alpha(\mu)}{4\epsilon}\right)^\epsilon \approx 1 - \frac{\alpha(\mu)}{4} \,. \tag{B40}$$

Inserting this into our expression for the non-interacting conductivity and taking the limit $\epsilon \to 0$, we obtain the following:

$$\sigma_0(\omega) = e_0^2 \frac{N}{16} \left(1 - \frac{\alpha(\omega)}{4}\right) \,, \tag{B41}$$

a modification of the bare-bubble result arising from the velocity renormalization. Remarkably, as found by TK, this term comines with the $\mathcal{O}(\alpha)$ diagrams that are also different in $d$ dimensions. Indeed, the vertex- and the self-energy diagrams yield, within the DR scheme, an interaction correction coefficient[6]

$$\mathcal{C}' = \frac{22 - 6\pi}{12} \,, \tag{B42}$$

as we have discussed, although this apparently leads to a result that disagrees with other regularization schemes, regularizing the full theory leads to the additional contribution in Eq. (B41). Including all contributions to order $\alpha$ gives:

$$\sigma(\omega) = \sigma_0(\omega) \left(1 + \mathcal{C}'\alpha(\omega) - \frac{\alpha(\omega)}{4}\right) \tag{B43}$$

$$= \sigma_0(\omega) \left(1 + \mathcal{C}\alpha(\omega)\right) \,, \tag{B44}$$

where

$$\mathcal{C} = \frac{19 - 6\pi}{12} \,. \tag{B45}$$

### 2. Wilson momentum-shell RG and Dimensional Regularization

In this section we will show that combining the Wilson momentum-shell RG (WRG) and the dimensional regularization (DR) yields a UV quirk. We obtain a non-commuting order of limits between the UV-cutoff $\Lambda$ introduced by the WRG and the parameter $\epsilon$ of the DR. In practice, this means that the additional renormalization factor appearing in the bare bubble term of the TK calculation [Eq. (B40)] is not present when we use the WRG in $d$ dimensions. This factor only occurs in the absence of any cutoff (pure DR), where it combines with the coefficient $\mathcal{C}'$ arising from the leading order interaction-correction diagrams (as discussed in the preceding section). In contrast, the WRG approach in $d$ dimensions directly obtains $\mathcal{C}$ without the factor Eq. (B40), as long as $\Lambda$ is sent to $\infty$ only at the end of the calculation.

#### a. Conductivity via Mishchenko's approach

Here, we evaluate the conductivity following Mishchenko's approach using the density-density correlator. Since the divergence is only present in the self energy diagram, it is straightforward to see that the impact on this divergence

is qualitatively different if we use dimensional regularization vs. a sharp cutoff. Within Mishchenko's approach, the conductivity is

$$\sigma(\omega) = \text{Lim}_{\mathbf{q}\to 0} \frac{\omega}{q^2} \text{Im} \chi_{00}(\mathbf{q}, \omega),$$
(B46)

with $\chi_{00}$ the retarded density-density correlator. As we have shown previously,[4] the self energy is

$$\Sigma(\mathbf{p}) = -e^2 \int_{P'} V(\mathbf{p} - \mathbf{p}') G(P'),$$
(B47)

$$= \frac{1}{4} \alpha v \mathbf{p} \cdot \boldsymbol{\sigma} \ln\left[\frac{4\Lambda c}{p}\right],$$
(B48)

with $c = e^{1/2}$. With this result, the contribution due to the self-energy type diagrams is (note we multiplied by 2 for the two diagrams):

$$\chi_{00}^{(1)}(Q)\big|_{bc} = -\frac{1}{2} N e^2 \alpha \int \frac{d^2 p}{(2\pi)^2} \int \frac{d\omega}{2\pi} \ln \frac{4\Lambda c}{p} \text{Tr}\left[\frac{-i\omega\sigma_0 - \mathbf{p}\cdot\boldsymbol{\sigma}}{\omega^2 + p^2} \frac{-i(\omega+\Omega)\sigma_0 - (\mathbf{p}+\mathbf{q})\cdot\boldsymbol{\sigma}}{(\omega+\Omega)^2 + (\mathbf{p}+\mathbf{q})^2} \frac{-i\omega\sigma_0 - \mathbf{p}\cdot\boldsymbol{\sigma}}{\omega^2 + p^2} \mathbf{p}\cdot\boldsymbol{\sigma}\right],$$
(B49)

where $Q = (\Omega, \mathbf{q})$. Evaluating the trace yields

$$\chi_{00}^{(1)}(Q)\big|_{bc} = -\frac{1}{2} N e^2 \alpha \int \frac{d^2 p}{(2\pi)^2} \int \frac{d\omega}{2\pi} \ln \frac{4\Lambda c}{p} \frac{-2}{(\omega^2 + p^2)^2} \frac{p^4 - 3\omega^2 p^2 - 2\omega\Omega p^2 - \mathbf{q}\cdot\mathbf{p}(\omega^2 - p^2)}{(\omega+\Omega)^2 + (\mathbf{p}+\mathbf{q})^2}.$$
(B50)

Expanding to quadratic order in $q$ and integrating over angles gives (keeping only the $q^2$ term)

$$\chi_{00}^{(1)}(Q)\big|_{bc} = -\frac{1}{2} N e^2 \alpha \frac{q^2}{8\pi^3} \int_0^\infty p \, dp \int_{-\infty}^\infty d\omega \ln \frac{4\Lambda c}{p} \frac{8p^2\pi}{(\omega^2 + p^2)^2} \frac{(\Omega+\omega)(\Omega+2\omega)[p^2 - \omega(\Omega+\omega)]}{[p^2 + (\omega+\Omega)^2]^3}.$$
(B51)

To proceed, we rescale $p \to \Omega p$ and evaluate the frequency integral. Thus we have

$$\chi_{00}^{(1)}(Q)\big|_{bc} = -\frac{1}{2} N e^2 \alpha \frac{q^2}{8\pi^3} \frac{1}{\Omega} \int_0^\infty p \, dp \ln \frac{v\Lambda c}{p\Omega} \frac{\pi^2(4p^2-1)}{p(4p^2+1)^2},$$
(B52)

$$= \frac{q^2 N\alpha}{64\Omega}.$$
(B53)

Note that the integral over $p$ was a sum of two terms due to the formula

$$\ln \frac{v\Lambda c}{p\Omega} = \ln \frac{v\Lambda c}{\Omega} - \ln p,$$
(B54)

However, the first such integral *vanishes*. Upon analytically continuing, we find the contribution to the conductivity

$$\sigma_{bc} = \sigma_0 \frac{\alpha}{4},$$
(B55)

which agrees with Eq.(13) of Mishchenko[2]

As we have noted, within this approach, the vertex diagram (also called the $d$ diagram) has no divergences, and so we can simply take Mishchenko's result for this diagram. The result is:

$$\sigma_d = \sigma_0 \alpha \frac{8 - 3\pi}{6},$$
(B56)

finally leading to:

$$\sigma = \sigma_0 \left(1 + \alpha \frac{19 - 6\pi}{12}\right),$$
(B57)

the expected result. Our next task is to see how dimensional regularization can give a different result.

### b. Dimensional regularization

We now consider how these results would change if we had instead worked in $d = 2 - \epsilon$ spatial dimensions, using spatial dimensionality to regularize the integral before taking the limit of $\epsilon \to 0$ at the end of the day. All that changes is that we would simply replace the logarithm coming from the self energy with the result from dimensional regularization, and also alter the dimensionality of the momentum integral. The self energy in $d$ dimensions is proportional to:

$$\Sigma(p) \propto p^{-\epsilon} \frac{\Gamma\left[\frac{1-\epsilon}{2}\right]\Gamma\left[\frac{3-\epsilon}{2}\right]\Gamma\left[\frac{\epsilon}{2}\right]}{\pi\Gamma[2-\epsilon]}, \tag{B58}$$

$$\simeq \frac{1}{\epsilon} - \frac{1}{2}\gamma + \ln 4 - \ln p. \tag{B59}$$

Now, as noted above, only the $\ln p$ part of the integral contributes. Given this fact, one may ask how we could get a different result in dimensional regularization since all that changes is the replacement $\ln \Lambda \to \frac{1}{\epsilon}$ (up to additional constant terms). In strictly two dimensions, but with the self energy evaluated in $d = 2 - \epsilon$ dimensions, the relevant integral is:

$$\int_0^\infty p\,dp\left[\frac{1}{\epsilon} - \frac{1}{2}\gamma + \ln\frac{4}{p}\right]\frac{4p^2 - 1}{p(4p^2 + 1)^2} = \int_0^\infty p\,dp\ln\frac{1}{p}\frac{4p^2 - 1}{p(4p^2 + 1)^2} = -\frac{\pi}{4}, \tag{B60}$$

where to get to the middle expression we used the fact that the contribution to the integral from the constant ($p$-independent) piece in square brackets vanishes. Clearly, this would yield the result Eq. (B57) whether the cutoff comes from $\Lambda$ or dimensional regularization. The difference is that, in $d$ spatial dimensions, the integration measure also changes. Therefore, if we change our integration to be in $d = 2 - \epsilon$ spatial dimensions, we obtain:

$$\int_0^\infty p^{1-\epsilon}dp\left[\frac{1}{\epsilon} - \frac{1}{2}\gamma + \ln\frac{4}{p}\right]\frac{4p^2 - 1}{p(4p^2 + 1)^2} = -\frac{\pi}{2}, \tag{B61}$$

where we took $\epsilon \to 0$ at the end of the calculation. Now, the part of the integral coming from the momentum-independent parts of the square brackets does not vanish, but is instead proportional to $\epsilon$ and yielding a finite contribution when multiplied by $\frac{1}{\epsilon}$. This small difference is seen to double the size of the $bc$ diagrams. Since the $d$ diagram does not change (since it is convergent) it would lead to the final conductivity correction proportional to $\mathcal{C}' = \frac{22-6\pi}{12}$. In Eq. (B61), we approximated the self energy by its $\epsilon \to 0$ limit. However, the same result holds if we instead used the full power-law expression Eq. (B58).

### c. Spatial dimension $d = 2 - \epsilon$ but sharp cutoff

We can also consider working in $d = 2 - \epsilon$ dimensions but with maintaining a UV cutoff $\Lambda$. This is what we are implicitly doing when we do momentum shell (Wilson) RG in $d$ dimensions. We expect, generally, that the self energy in this case will be of the form:

$$\Sigma(p) \propto \int_p^\Lambda \frac{q^{d-1}dq}{q^2} = \frac{1}{\epsilon}\left(p^{-\epsilon} - \Lambda^{-\epsilon}\right). \tag{B62}$$

Plugging this into our integral, we obtain:

$$\int_0^\infty p^{1-\epsilon}dp\frac{1}{\epsilon}\left(p^{-\epsilon} - \Lambda^{-\epsilon}\right)\frac{4(vp)^2 - \omega^2}{p(4(vp)^2 + \omega^2)^2} = \left(\frac{\omega}{v}\right)^{1-2\epsilon}\frac{\frac{\pi}{4}\left(\frac{\omega}{v\Lambda}\right)^\epsilon - \frac{\pi}{2}}{\omega^2}, \tag{B63}$$

which is the UV quirk. The order of limits of the UV-cutoff $\Lambda$ and the dimensional parameter $\epsilon$ do not commute. If we first took the limit of $\epsilon \to 0$ and at the end of the calculation the limit $\Lambda \to \infty$, we would obtain $\mathcal{C}$, whereas if we took first $\Lambda \to \infty$ and at than $\epsilon \to 0$, we would have $\mathcal{C}'$.

$$\lim_{\epsilon \to 0}\left(\lim_{\Lambda \to \infty}\left[C\left(\epsilon, \frac{\omega}{\Lambda}\right) = \frac{22 - 6\pi - 3\left(\frac{\omega}{v\Lambda}\right)^\epsilon}{12}\right]\right) = \mathcal{C}' \tag{B64}$$

$$\lim_{\Lambda \to \infty}\left(\lim_{\epsilon \to 0}\left[C\left(\epsilon, \frac{\omega}{\Lambda}\right) = \frac{22 - 6\pi - 3\left(\frac{\omega}{v\Lambda}\right)^\epsilon}{12}\right]\right) = \mathcal{C}. \tag{B65}$$

Within WRG, the solution to this ultraviolet quirk is to always maintain nonzero $\Lambda$, first setting $\epsilon \to 0$, a procedure which unambiguously yields the coefficient $\mathcal{C}$. If we set $\Lambda \to 0$ first then, although the theory is regularized, additional singularities appear in the limit $\epsilon \to 0$. A correct handling of this limit requires regularizing the full theory (not just the diagrams) as reviewed in the preceding subsection (and again finally yielding $\mathcal{C}$ for the interaction correction coefficient).

---



\end{document}